\documentclass[12pt]{article}
\usepackage[utf8]{inputenc}
\usepackage{mathtools}
\usepackage[english]{babel}
\usepackage[T1]{fontenc}
\usepackage[font={small,it}]{caption}
\usepackage[a4paper,top=3cm,bottom=2cm,left=3cm,right=3cm,marginparwidth=1.75cm]{geometry}
\usepackage{adjustbox}
\usepackage{graphicx}
\usepackage{rotating}
\usepackage[font=footnotesize,textfont=it]{caption}
\usepackage{xcolor}
\usepackage{soul}
\usepackage{tabularx}
\graphicspath{ {icons/} } 
\usepackage{titlesec}

\usepackage{amsmath}
\usepackage[colorlinks=true,allcolors=blue]{hyperref}
\usepackage{wrapfig}

\title{\large \bf The Knowledge Mobility of Renewable Energy Technology}
\author{\normalsize Persoon, P.G.J.\\
\texttt{\normalsize P.G.J.Persoon@tue.nl}
  \and
   \normalsize Bekkers, R.N.A.\\
  \texttt{\normalsize R.N.A.Bekkers@tue.nl}
  \and
  \normalsize Alkemade, F.\\
  \texttt{\normalsize F.Alkemade@tue.nl}
}
\date{\normalsize March 2021}

\begin{document}

\maketitle

\begin{abstract}
In the race to achieve climate goals, many governments and organisations are encouraging the local development of Renewable Energy Technology (RET). The spatial innovation dynamics of development of a technology partly depends on the characteristics of the knowledge base on which this technology builds, in particular the analyticity and cumulativeness of knowledge. Theoretically, greater analyticity and lesser cumulativeness are positively associated with more widespread development. In this study we first empirically evaluate these relations for general technology and then systematically determine the knowledge base characteristics for a set of 14 different RETs. We find that, while several RETs (photovoltaics, fuel-cells, energy storage) have a highly analytic knowledge base and develop more widespread, there are also important RETs (wind turbines, solar thermal, geothermal and hydro energy) for which the knowledge base is less analytic and which develop less widespread. Likewise, the technological cumulativeness tends to be lower for the former than for the latter group. This calls for regional and country level policies to be specific for different RETs, taking for a given RET into account both the type of knowledge it builds on as well as the local presence of this knowledge.
\end{abstract}

\section{Introduction} 
 The widespread development and use of Renewable Energy Technologies (RETs) is an essential part of the transition towards a carbon-free society \cite{ipcc_climate_2014}. The ability of a country or region to participate in the development of a technology not only depends on the locally available knowledge and capabilities \cite{li_emergence_2020}, but also on the characteristics of {\it the knowledge base of that technology}. More specifically, Binz and Truffer \cite{binz_global_2017} argue that it is typically easier to enter in knowledge fields with a more global (and 'footloose') knowledge base when compared to knowledge bases that are more local (and 'sticky'). These characteristics of the knowledge base have been linked to different modes of knowledge production; global, footloose knowledge to a 'Science-Technology and Innovation (STI) mode' observed in science-based industries that lean very much on analytical knowledge, and local, sticky knowledge bases to a 'Doing, Using and Interacting (DUI) mode' observed in engineering-based industries that lean very much on synthetic knowledge \cite{jensen_forms_2007, asheim_regional_2016}. RETs may thus differ in the extent to which their development spreads over countries or regions, i.e., the mobility of their knowledge base. Where the development of some RETs may take place in STI-mode, widespread and expanding, the development of other RETs may take place in DUI-mode, concentrated and difficult to relocate. This has implications for countries that seek to move closer to the knowledge frontier through technology and R\&D investments as this may be easier for more footloose technologies \cite{keller_international_2004}. Understanding the knowledge base characteristics of renewable energy technologies---in particular the knowledge dimensions relating to the spatial dynamics of innovation---is thus pivotal input for targeted and evidence-based renewable energy policies.
 
 Earlier studies analyzing RETs as a single technology class find that RETs on average build more on analytical and geographically distant knowledge than other technologies \cite{ocampo-corrales_knowledge_2020}, and that they benefit greatly from knowledge flows transcending national borders \cite{noailly_multinational_2015, garrone_innovation_2014}. However, recent studies at the more detailed level of individual technologies find considerable heterogeneity in the extent to which RETs build on analytical knowledge \cite{persoon_science_2020, hotte_rise_2020}. For example, the science dependence of some RETs, such as wind turbines, is relatively low, and closer to fossil fuel based energy technologies, whereas photovoltaics and non-fossil fuels are characterized by a high science dependence. Similar variations have been observed in other dimensions of the knowledge base that may affect the place-dependence of RETs such as the cumulativeness \cite{persoon_how_2021}, which is associated with greater geographical concentration of innovative activities \cite{malerba_persistence_1997,breschi_technological_2000}. Building on the framework outlined by Binz and Truffer \cite{binz_global_2017}, we systematically investigate these different characteristics of the knowledge base of RETs in order to assess whether these technological are more local or global in nature. More specifically, we map analyticity, cumulativeness and knowledge mobility for the knowledge base of 14 different RETs.

 The remainder of this paper is structured as follows. In Section \ref{theory} we discuss the theoretical background of the mentioned knowledge dimensions and our expectations for the different RETs. Then in Section \ref{methodology} we explain how we measure the knowledge dimensions and distinguish the different RETs. Subsequently, we report our main observations in Section \ref{results}, discuss some deeper implications and shortcomings in Section \ref{discussion} and end with a number of conclusions and policy recommendations in Section \ref{conclusions}. 
 
 \section{Theory}\label{theory}

The process of knowledge creation and innovation in a certain technology depends for an important part on characteristics of the body of knowledge on which this technology builds\cite{asheim_knowledge_2005, breschi_technological_2000}, henceforth referred to as the 'knowledge base' of a technology. In the following we will discuss three dimensions of the knowledge base which can theoretically be linked to spatial dynamics of innovation: the analyticity (Section \ref{analyticity}) the cumulativeness (Section \ref{cumulativeness}), and the knowledge mobility (Section \ref{mobility}). We then discuss our expectations for these dimensions for the different RETs (Section \ref{expectations}).

\subsection{Analyticity of knowledge}\label{analyticity}

Knowledge bases are described to consist of three types of knowledge: analytic, synthetic and symbolic knowledge \cite{asheim_knowledge_2005,moodysson_explaining_2008}. In this context, analytic knowledge is understood to be science-based, created in deductive processes based on formal models. Synthetic knowledge is understood as engineering-based, created through the application of existing knowledge or through inductively combining existing knowledge. Finally, symbolic knowledge is characterized as cultural or artistic knowledge.\footnote{In this research we will mostly focus on the distinction between analytic and synthetic knowledge.} As the knowledge base of technologies often contains multiple types of knowledge, it is more instructive to think about the extent to which it is analytic as a spectrum or a scale. In this line of thinking, we define the analyticity of a knowledge base as the extent to which it consists of analytic knowledge. 

The analyticity of knowledge has been associated with several other theoretical dimensions of knowledge. First, where analytical knowledge is associated naturally with basic research, i.e. research aimed at truth-finding, synthetic knowledge is associated with applied research, i.e research aimed at solving practical problems \cite{bentley_relationship_2015,oecd_frascati_2015}. Technologies that strongly depend on analytic knowledge are therefore understood to have stronger ties with the natural sciences. While closely related, basic and analytic (or applied and synthetic) cannot be considered synonyms: basic research may occasionally produce synthetic knowledge, and vice versa. Second, and closely related, where analytic knowledge is universal and theoretical, synthetic knowledge is context specific and practice related \cite{moodysson_explaining_2008}. It is therefore expected that it is more difficult to work with synthetic knowledge outside the context in which it was developed, that is synthetic knowledge is stickier and place dependent. Third, analytic knowledge is often associated with codified knowledge and synthetic knowledge with tacit knowledge. However, here too, there are certainly exceptions. Not all published work is easy to fully understand or reproduce without the aid of those that produced the work. Authors have therefore argued that there sometimes is a tacit element to analytic knowledge as well \cite{moodysson_explaining_2008}.  

\subsection{Technological cumulativeness}\label{cumulativeness}

Knowledge bases can also be characterized by their 'technological cumulativeness', the idea that today's technologies are developed by building on the insights from yesterday's technologies and will themselves be used to develop the technologies of tomorrow \cite{trajtenberg_university_1997, breschi_technological_2000}.  Perspectives on the exact meaning of technological cumulativeness however vary, for an overview of this discussion we refer to \cite{persoon_how_2021}. In this work, we understand a technological development to be cumulative when a later technological result depends on an earlier technological result. In the context of technological knowledge, we broadly interpret this dependency as the usage, modification or improvement of earlier ideas. In this line of thinking, we understand the knowledge base of a technology to be more cumulative when the developments in this technology are more cumulative. This allows us to define the cumulativeness of a technology as the extent to which developments within this technology are cumulative. In other words, the more a technology builds on its earlier developments, the greater its cumulativeness. 

Technological cumulativeness is often mentioned as a defining characteristic of a \textit{technological regime}, which is a description of the relevant environment or circumstances for companies and organizations to innovate \cite{nelson_search_1977, breschi_technological_2000}. When a technological regime is characterized by high cumulativeness, established parties largely dominate innovative activities and it is relatively hard for new parties to enter. Highly cumulative technologies allow firms or organisations to gain absorptive capacity through learning and specialization \cite{cohen_absorptive_1990}. Within a technological regime therefore, greater cumulativeness is associated with a greater appropriability of innovation and a greater geographical concentration of innovative activities \cite{malerba_schumpeterian_1996, malerba_persistence_1997}. 

In an earlier contribution, where the cumulativeness was explicitly approached as the extent to which a technology builds on its earlier developments, we established that the cumulativeness of a technology increases approximately linearly with the size of its knowledge base, at a technology specific rate \cite{persoon_how_2021}. Especially when cumulativeness is compared across technologies, this suggests that next to considering the cumulativeness of a technology, it will be useful to consider the rate at which the cumulativeness increases, i.e., the cumulativeness relative to the size of the knowledge base.  

When cumulative developments stretch over longer periods of time, the products associated with a technology tend to become 'more complex', meaning that the number of interrelated (functional) parts of a product architecture increase. Technological complexity is therefore often associated with greater cumulativeness. The complexity of technologies is in the literature however mostly approached anecdotally or on a case to case basis, as there is no general agreement on a single objective measure for complexity \cite{vaesen_complexity_2017}. 

\subsection{Knowledge mobility}\label{mobility}

Where some types of knowledge travel easily from one place to another, other types are bound to a certain location. In order to investigate this dimension of knowledge, we define the knowledge mobility as the extent to which knowledge travels geographically. By geographical traveling, we mean that knowledge developed in one location is subsequently used or applied in another location, where the two locations are separated by a geographical distance. High knowledge mobility then corresponds to knowledge that travels with ease to more distant locations, i.e., 'footloose knowledge'. Low knowledge mobility corresponds to Knowledge that travels less easily, i.e. 'sticky knowledge'. A highly mobile body of knowledge is thus expected to travel farther, in other words, we expect mobile knowledge to be more widespread (or less concentrated) than sticky knowledge.

Knowledge bases characterized by greater analyticity are expected to be more mobile (or 'footloose') \cite{asheim_constructing_2011, herstad_industrial_2014}. A motivation for this first expectation is the universality and theoretical nature of analytic knowledge, which almost per definition implies time, location and application independence. The context specificity and practical nature of synthetic knowledge on the contrary makes it more time, location and application bound. Another motivation is the supposed association with codified knowledge: what is written down travels easier than know-how fixed in the minds of experts \cite{lundvall_learning_1994,gertler_tacit_2003}. As mentioned earlier though, this association is also criticized. These motivations also count when the causality is reversed: when innovative activities are fixed and concentrated geographically, there may be less need to formalize or rationalize findings because knowledge is communicated orally, developed during collaboration and hence may remain largely tacit and fragmented. A gradual shift towards knowledge more synthetic in nature is thus expected when engineers work close together. Likewise, when collaborators are forced to communicate their results at a distance, it may stimulate them to formalize or rationalize their implicit ideas or intuitions. 

Knowledge bases characterized by higher cumulativeness are expected to be stickier\cite{herstad_industrial_2014}. A motivation for this second expectation is the expected greater geographical concentration of innovative activities in technological regimes characterized by high cumulativeness. With greater geographical concentration, we expect the development to be less widespread and hence to travel shorter distances. Note that this relation too can be reversed, namely that the knowledge is concentrated because it is sticky. Another motivation for this second expectation comes from the association between cumulativeness and technological complexity: technologically complex knowledge does not travel well \cite{balland_geography_2017}. Working with or improving a complex system from a distance is challenging, because it becomes more difficult to experiment or interact with the system. 

\subsection{Knowledge dimensions of RETs}\label{expectations}
In this research we aim to investigate how the knowledge dimensions vary for different Renewable Energy Technologies (RETs). While a 'technology' can be approached or characterized from many different angles, we will in this contribution largely focus on the knowledge properties of technologies, hence approaching the different RETs as distinct bodies of knowledge. The knowledge properties may cover various aspects of the technology, for example, how the technology operates or how it is constructed. While the purpose of the various RETs largely coincide (enable the generation of renewable energy), the renewable energy sources that the various RETs exploit (and thus their working principles) fundamentally differ. Following the International Renewable Energy Agency (IRENA), we distinguish between geothermal, hydropower, ocean, wind, solar thermal, solar photovoltaic, and bio-energy \cite{irena_global_2018}. In addition, we include a number of enabling technologies allowing for the storage of energy such as hydrogen technology, and three energy related technologies that are not entirely renewable yet may help reduce CO2 emissions: nuclear energy, carbon capture \& storage (ccs) and clean combustion. We will provide a more precise list of the individual RETs in the next section. 

Earlier contributions have indicated that RETs generally build strongly on scientific knowledge, suggesting a highly analytic knowledge base \cite{ocampo-corrales_knowledge_2020}. In agreement with this finding, innovative activities related to RETs are observed to take place on ever larger geographic scales \cite{noailly_multinational_2015, garrone_innovation_2014}. At same time the knowledge bases are known to vary greatly across different RETs \cite{barbieri_knowledge_2020} and across energy technology in general \cite{nemet_inter-technology_2012}. More specifically, we know there is a large variation across RETs in the extent to which the knowledge base is science based \cite{persoon_science_2020, hotte_rise_2020}. Where photovoltaics, non-fossil fuels and to some extent fuel-cells and hydrogen technology were found to be more science based, wind turbines, hydroelectric and geothermal energy were found to be less science based. The more a RET depends on science, the more analytic its knowledge base, the greater a knowledge mobility we expect for these technologies. 

While the development of different RETs has been studied in numerous contributions, it appears that the current literature lacks a systematic comparison of the cumulativeness across different RETs. Even though the size of the knowledge base varies across RETs, this does not automatically translate to a similar variation in cumulativeness \cite{persoon_how_2021}. The closely related technological complexity however does appear to vary largely across RETs. Interpreting a larger technological complexity for systems with many interdependent parts, RETs such as wind turbines, geothermal energy, nuclear fission and energy from sea are identified as rather complex \cite{huenteler_how_2016}, more complex than photovoltaics and non-fossil fuels.\footnote{In some cases the technological complexity varies with different applications of a technology. For example for solar thermal energy, the systems in domestic use are limited to elements that efficiently capture and store heat, and are therefore relatively simple, whereas the systems used in power plants are typically larger, contain more different elements and have the additional features of concentrating the heat and  transforming it to electric power, making these systems far more complex.} The variation in technological complexity suggest there may be large variation across RETs in cumulativeness too (though this needs empirical validation). As the knowledge bases characterized by high cumulativeness tend to be stickier, we expect the higher cumulativeness RETs to show a lower knowledge mobility. Taking a slightly different perspective, Binz, Tang and Huenteler distinguishes between 'complex engineered systems for specialized users' and 'standardized mass-manufactured goods', wind-turbines again being an example of the former and household energy storage systems, stationary fuel-cells and photovoltaics an example of the latter \cite{binz_spatial_2017}. Based on their findings about photovoltaics, they expect life-cycle dynamics of the latter group to be more 'spatially fluid' than the former.  

Summarizing, we expect to observe a greater knowledge mobility for RETs with a stronger dependence on science {\it and} RETs characterized by lower cumulativeness. 

\section{Methodology} \label{methodology} 

This section presents the methods used to measure analyticity, cumulativeness and knowledge mobility for RETs. First, we discuss our data and present indicators for the knowledge dimensions. Subsequently we discuss our selection of various RETs and present some descriptive statistics.  

\subsection{Patents}

Earlier approaches to measuring the analytic-synthetic knowledge distinction were often based on data from questionnaires or professional occupations \cite{moodysson_explaining_2008, plum_analysing_2012, martin_measuring_2012}. While useful, these data are largely an \textit{indirect} measure of knowledge characteristics, because they are based on the characteristics of the people that use or produce the knowledge, instead of the knowledge itself. In this contribution we aim to \textit{directly} measure the knowledge characteristics by studying codified forms of knowledge, more specifically, patent data. 

Patent data directly represent technological knowledge, containing a wealth of detailed information about both the technological content as well as the inventor or applicant. Furthermore, the citations in patents, both to other patents and scientific literature, to some extent allow us to proxy knowledge connections and flow. While patent data offer a unique opportunity to quantitatively study novel and relevant technological knowledge development, there are also some limitations. Not all technology is patented and not all patents represent relevant technological developments. While acknowledging these disadvantages, we believe that for the purpose of understanding RET development there is a great potential for patent data. 

A possible criticism of the usage of patent data to proxy the analytic-synthetic distinction is the supposed association with the codified-tacit distinction: as patents are codified knowledge, we risk observing analytic knowledge only. However, as mentioned earlier, the association with codified-tacit distinction is also criticized, and we strongly believe that patents, a key element of engineering practices, may equally well contain a large degree of synthetic knowledge. Our approach is therefore, that within codified knowledge, there may be different degrees of analytic knowledge. More specifically in the context of technological knowledge, the more a body of codified knowledge can be associated with scientific activity, the greater we will interpret its degree of analytic knowledge.    

Finally, we shortly comment on the geography of patents. In this research we will do a separate analysis for patents from the EPO (European Patent Office) and the USPTO (United States Patent and Trademark Office), henceforth 'EP patents' and 'US patents' respectively. There are two reasons for this choice. First, different patent offices, but in particular EPO and USPTO, have institutionalized different rules for citation, hence limiting the analysis of knowledge connections to one patent office may give biased results. Second, an applicant files a patent with an office if there is market potential in the geographical jurisdiction of that office. As we are interested in the worldwide geography of innovation, we do not want to limit the analysis to a single geographical jurisdiction.   

\subsection{Indicators}\label{indicators}

For the analysis of analyticity we will mostly use the scientific character of this type of knowledge. To proxy for a given technology the dependence on science and the scientific content of the knowledge base we define the following indicators: 
\begin{itemize}
    \item The \textit{science dependence}(sd) of a technology is defined as the average number of references to scientific literature per patent. A reference in a patent to a scientific source can be interpreted as a dependency link, suggesting that scientific knowledge was somehow relevant in the content of the patent. The more scientific sources a patent therefore refers to, the more we expect it to be science based. We therefore take the science dependence as an indicator of analytic knowledge. 
    \item The \textit{science dependence fraction}(sdf) of a patent is defined as its number of references to scientific literature  divided by its total number of references. To obtain the sdf of a technology, we take the sdf of each patent in that technology and take the average. Hence where the sd is based on the absolute number of references, the sdf is based on the relative number of references, thus taking into account variation across patents and technologies in the number of references. A similar indicator was earlier used in \cite{hotte_knowledge_2021, hotte_rise_2020}.
    \item The \textit{university fraction}(uf) of a technology is defined as the number patents in that technology for which the inventor or applicant is university\footnote{As we will see later it more correct to speak of university-related organisations} affiliated divided by the total number of patents in that technology. When the inventor is affiliated with a university, we expect the patent to be based more on scientific knowledge than the average patent from non-scientific organisations. We therefore take the university fraction to be an indicator of analytic knowledge.
    \end{itemize}
To proxy the cumulativeness we will use 
\begin{itemize}
    \item The \textit{internal dependence}(id) of technology is defined as the average number of internal references per patent. An internal reference is a reference in a patent to a patent within the same technology, which can be interpreted as a dependency link from the technology to itself. Cumulativeness can be interpreted as the extent to which a technology builds on itself. This indicator was earlier used in \cite{persoon_how_2021}. For an approach based on general references, we refer to \cite{apa_knowledge_2018}.  
    \item The \textit{internal dependence fraction}(idf) of a patent is defined as its number of internal references divided by its total number of patent references\footnote{Alternatively, we can also include the references to scientific and/or other sources in this total. However, in this contribution we choose to define it as a fraction of patent references only, so that we can consider it to be independent from the sdf}. To obtain the idf of a technology, we take the idf of each patent in that technology and take the average. Hence where the id is based on the absolute number of references, the idf is based on the relative number of references, thus taking into account variation across patents and technologies in the number of references.
    \item The \textit{relative internal dependence}(rid) of a technology is defined as the internal dependence of that technology relative to its total number of patents, or equivalently, the number of internal references per \textit{patent squared}. As explained earlier, the internal dependence tends to increase linearly with the number of patents. When we compare technologies with different number of patents or when we are interested in the rate at which the cumulativeness increases, it is therefore useful to additionally consider the cumulativeness per patent.
\end{itemize}
 For the knowledge mobility we define the following indicators:
  \begin{itemize}
        \item The \textit{inter-patent distance}(ipd) of a technology is defined as the average geographic distance between each pair of patents within that technology. From the inventor or applicant addresses in patents we can create an overview of the approximate\footnote{That is approximately, as there is no guarantee the actual process of inventing took place at the mentioned address} locations of inventing. The mutual distances between patents can thus be used to proxy the geographical spread of inventing in a certain technology. 
        \item The \textit{reference distance}(rd) of a patent is defined as the average geographic distance between that patent and the (set of) patent(s) it refers to\footnote{We take the reference distance of a patent which does not refer to any other patent to be undefined}. The reference distance of a technology is defined as the average reference distance per patent. Where the inter-patent distance proxies the geographical spread, it does not directly proxy the possible knowledge flow between distant places. With the reference distance we therefore additionally consider the kilometers covered by references to obtain a better estimate of the actual movement of knowledge. Note however that the reference distance also includes references to other technologies, thus to some extent also measuring the knowledge flow of other technologies.\footnote{Excluding the references to other technologies can be demonstrated to result in an indicator very closely related to the inter-patent distance}
  \end{itemize} 
Note that all of these indicators can be determined for technologies (i.e. groups of patents) and a selection of these indicators can also be determined on the level of individual patents. In the first part of our analysis, we will use the indicators acting on the level of individual patents to establish a baseline and demonstrate more general relations between analyticity, cumulativeness and knowledge mobility (where the patents are not necessarily confined to the considered technologies).

This analysis is mainly based on data from Patstat (spring 2020 edition) focusing on European and US patents. As a consequence, there are number of subtleties involved with the actual measurement of the indicators: 
\begin{enumerate} 
\item We count as a 'patent' each unique DOCDB patent family, where an 'EP patent' represents each unique family with an EPO patent application and likewise for 'US patent' but then for USPTO applications\footnote{The Patstat records of USPTO applications are biased to granted patents before the year 2000. As our focus is not on the time development we expect this to be a minor issue for our purpose}.  
\item To identify the references to scientific literature we use the type 's' classification of the cited non-patent-literature (NPL), which signals articles in journals and periodicals. Where in Patstat the NPL appears to be classified rather well for EP patents, for the US patents the large majority of NPL, probably due to a lacking of rich structure in references, is classified in the general category 'a' (abstract of no specific kind). To obtain a better indication which fraction of the cited NPL is actually scientific, we use the database by Marx and Fuegi \cite{marx_reliance_2020} which links the references in patent applications to scientific publications, and is accurate for US patents. 
\item To identify the inventors or applicants affiliated with a university we use the automatized sector allocation in Patstat of persons \cite{magerman_data_2006, van_looy_data_2006}. This classification however allows an applicant to be allocated to multiple sectors. For the university fraction we include each patent where at least of one the allocations is the 'UNIVERSITY' sector. We therefore also include organisations closely related to the university, making it more correct to speak of 'university-related organisations'. 
\item To link the patents to geographical coordinates we use the 'Geocoding of worldwide patent data' database (shortly 'Geocoding') constructed by Rassenfosse, Kozak and Seliger \cite{de_rassenfosse_geocoding_2019} based on the applicant or inventor addresses. The Geocoding database is limited to first filed patent applications, which we linked back to patent families using Patstat. This research is based on the Geocoding table with inventor addresses. Yet, as the makers of the database acknowledge, disambiguation of inventors and applicants is generally challenging and a research task on its own. Indeed a quick comparison with the table bases on applicant addresses does not seem to amount to substantially different results. 
\item The addresses of inventors of EP patents are not necessarily confined to Europe, and likewise for US patents and the US. The typical reference distance of a EP patent with an inventor from the US however structurally differs from that with an inventor from Europe: the reference distance is location specific. While these variations are expected to average out when the number of patents in a technology is large, this effect may be disproportional for technologies with a smaller number of patents. To demonstrate this effect, we determine the reference distance from EP patents with US inventors and vice versa and compare these to the reference distances of EP (US) patents with European (US) inventors in Appendix \ref{app_2}. To account for this effect, when we determine the reference distance of EP patents we sub-select the patents with an inventor in Europe. Likewise when we determine the reference distance of US patents we sub-select the patents with and inventor from the US. These sub-selections contain for most of the technologies considered the majority of patents. Note that, while we sub-select patents based on the location of the inventor, the references in these may still be to patents from  inventors located anywhere in the world.
\end{enumerate}

\subsection{Technology selection and descriptive statistics}

We base our selection of energy generating technologies on the set of renewable energy sources identified by the International Renewable Energy Agency IRENA \cite{irena_global_2018}), including geothermal, hydropower, ocean, wind, solar and bioenergy. As mentioned earlier, these energy generating technologies are complemented with set of technologies relating to energy storage and a set of technologies which may not be considered fully renewable but nonetheless help reducing greenhouse gas emissions, such as nuclear energy, clean combustion and carbon capture and storage (ccs). For an overview see Table \ref{RETS}. To identify the patents associated with these (partial) RETs we use the Cooperative Patent Classification (CPC) used by both EPO and USPTO, or more specifically the CPC tagging scheme 'Y02' which identifies technologies with the potential to mitigate climate change \cite{veefkind_new_2012}.
Each of these RETs correspond then to a collection of patents classified on the group or subgroup level in CPC. The various technologies and corresponding CPC descriptions are shown in Table \ref{RETS}, including the symbols which represent them in later figures. Note that a substantial number of EP and US patents are members of the same patent family, hence there is a substantial overlap between both data-sets. 

\begin{table}
  \footnotesize
\begin{tabularx}{\linewidth}{|>{\centering\arraybackslash}m{2cm}|>{\centering\arraybackslash}m{4.5cm}|c|c|c|c|} 
\textbf{Technology} & \textbf{CPC description} & \textbf{CPC code} & \textbf{EP} & \textbf{US} & \textbf{in} \\
 &  &  & \textbf{patents} & \textbf{patents} & \textbf{common} \\
\hline
\includegraphics[width=20pt]{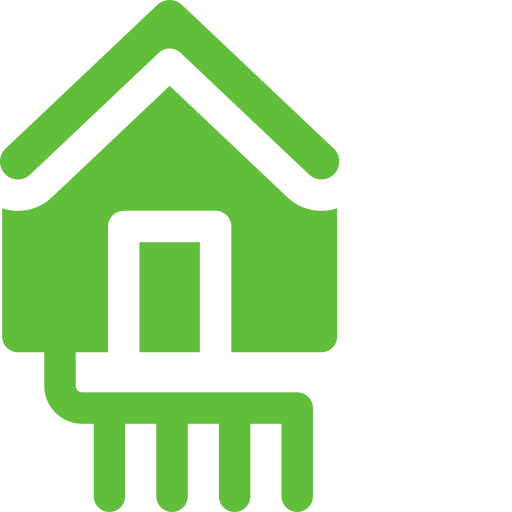} & Geothermal Energy & Y02E 10/1 & 495 & 1088 & 240 \\
  \hline
  \includegraphics[width=20pt]{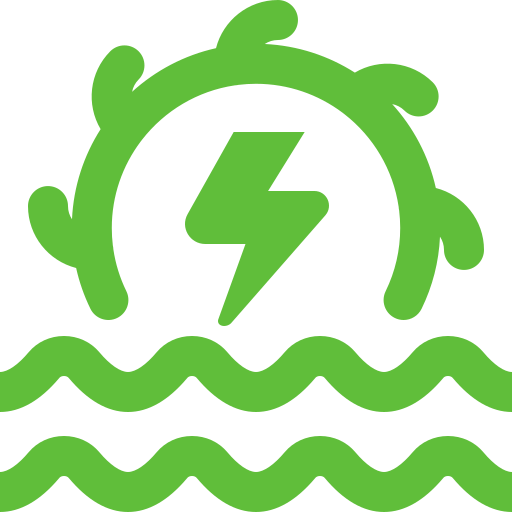} & 
    Hydro Energy & Y02E 10/2 & 1865 & 6223 & 1159 \\
    \hline
  \includegraphics[width=20pt]{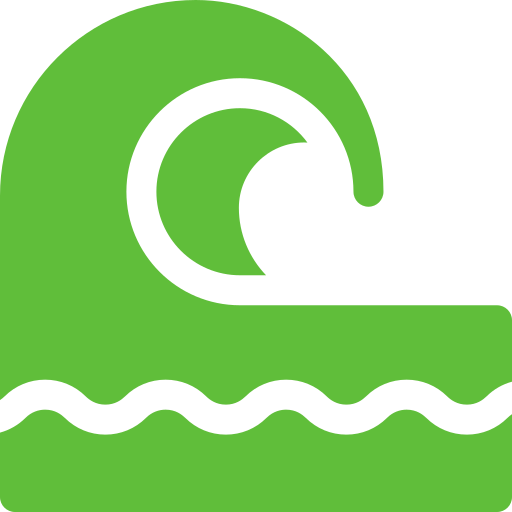} &  Energy from the sea, e.g. using wave energy or salinity gradient  & Y02E 10/3 & 1228 & 2624 & 902 \\
    \hline
  \includegraphics[width=20pt]{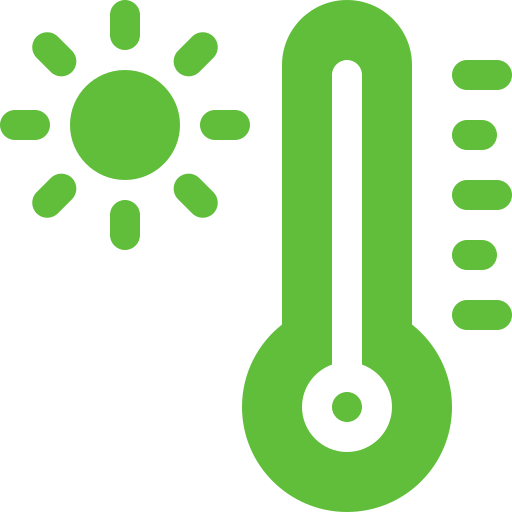} & Solar thermal energy, e.g. solar towers & Y02E 10/4 & 5425 & 11247 & 3034 \\
    \hline
  \includegraphics[width=20pt]{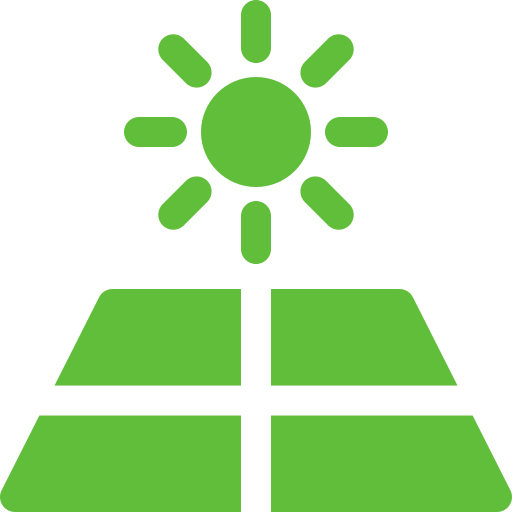} & Photovoltaic energy (photovoltaics) & Y02E 10/5 & 14947 & 31490 & 12492 \\
    \hline
   \includegraphics[width=20pt]{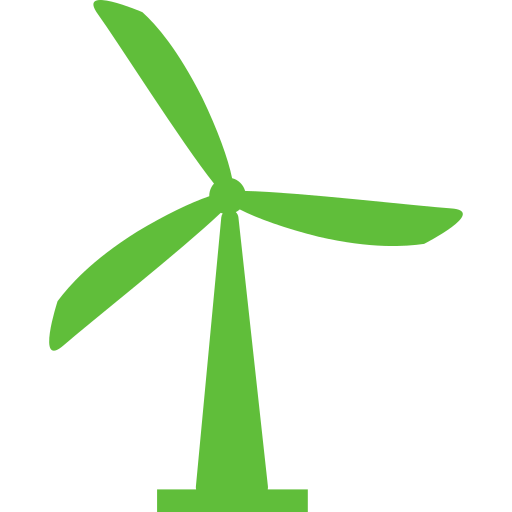} & Wind energy (wind turbines) & Y02E 10/7 & 10112 & 16454 & 7471 \\
    \hline
   \includegraphics[width=20pt]{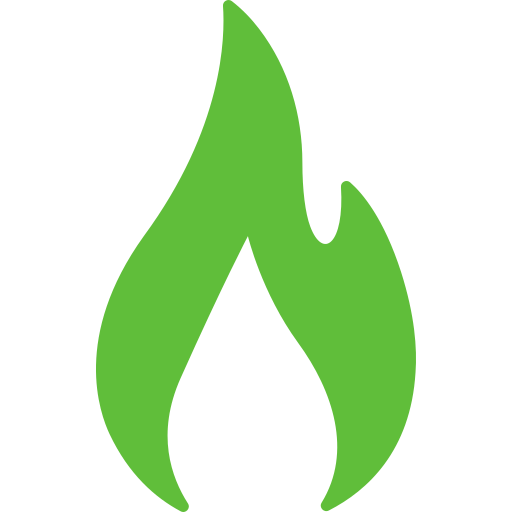} & Combustion technologies with mitigation potential (clean combustion) & Y02E 20 & 4956 & 7646 & 3575 \\
    \hline
   \includegraphics[width=20pt]{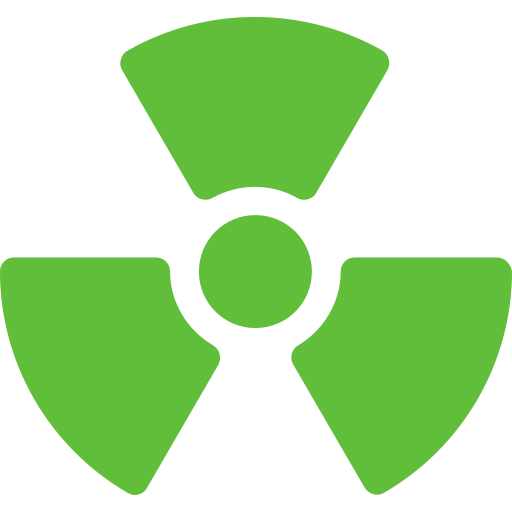} & Nuclear fission reactors & Y02E 30/3 & 1337 & 4325 & 1038 \\
    \hline
   \includegraphics[width=20pt]{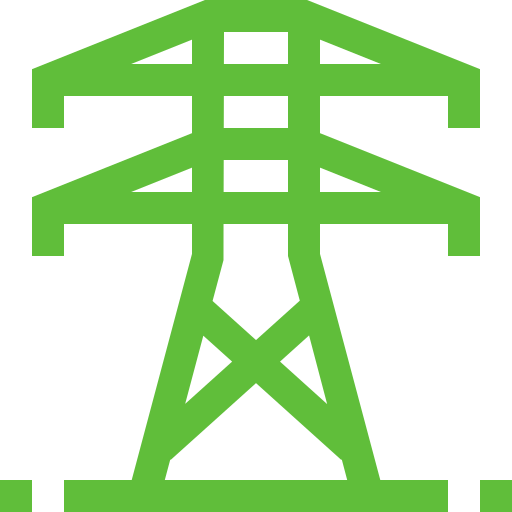} & Technologies for an efficient electrical power
generation, transmission or distribution (electric grids) & Y02E 40 & 2171 & 4031 & 1718\\
    \hline
   \includegraphics[width=20pt]{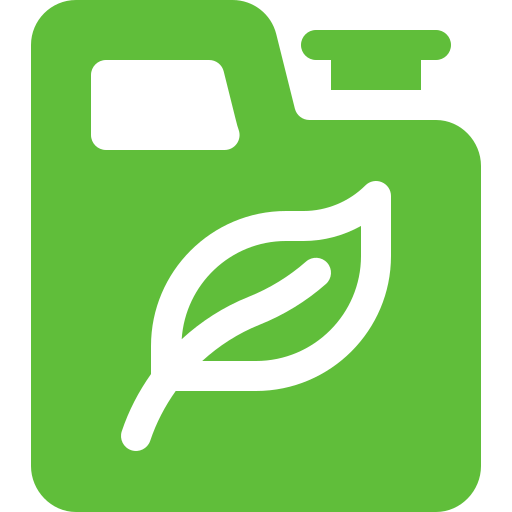} & Technologies for the production of fuel of nonfossil origin (non-fossil fuels) & Y02E 50 & 6310 & 9625 & 4548\\
    \hline
   \includegraphics[width=20pt]{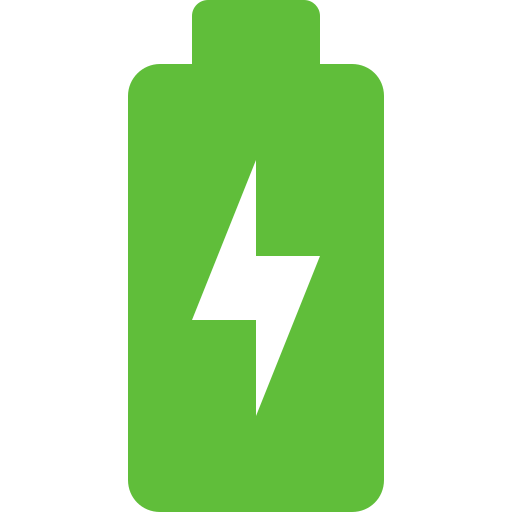} & Energy storage using batteries, capacitors, thermal or mechanical systems. & Y02E 60/1 & 8858 & 17502 & 7166 \\
    \hline
   \includegraphics[width=20pt]{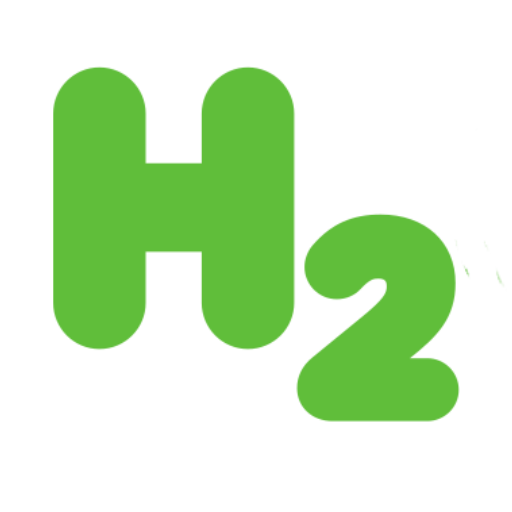} & Hydrogen Technology & Y02E 60/3 & 4029 & 7307 & 3220 \\
    \hline
    \includegraphics[width=20pt]{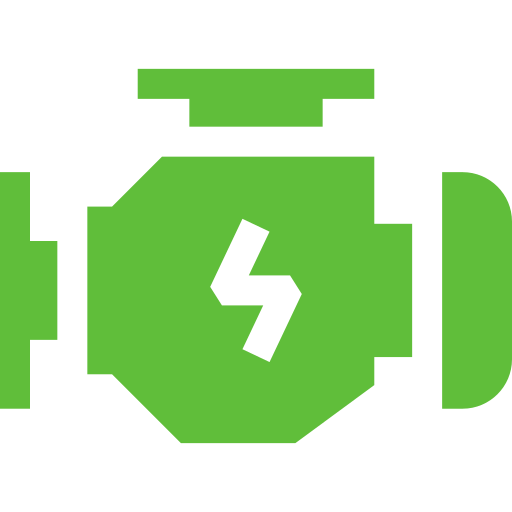} & Fuel cells & Y02E 60/5 & 3501 & 7254 & 3152 \\
    \hline
    \includegraphics[width=20pt]{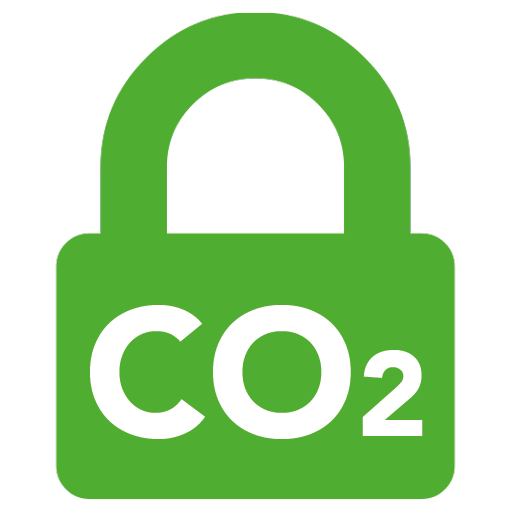} & Carbon capture and storage (ccs) & Y02C & 3791 & 6297 &  3091\\
    \hline
\end{tabularx}
\caption{\textbf{Symbols, CPC codes and total number of patents of selected RETs} To the CPC descriptions we added for some technologies a shortname (in brackets). The final column indicates the number of EP and US patents of which the family is the same.}
\label{RETS}
 \end{table}

In Table \ref{RETS_des} we include the descriptive statistics for a number of indicators discussed in the previous section. All of these indicators are only positive and characterized by distributions skewed towards the value zero, which is in line with the observation that the standard deviations are of the same order as the averages. The variation across technologies is substantial especially across the analyticity and cumulativeness indicators. The science dependence of non-fossil fuels is much higher than that of the other RETs. This is in line with earlier findings  \cite{persoon_science_2020, hotte_rise_2020} and may  be related to the strong link of non-fossil fuels to fields such as (Applied) Microbiology, Biochemistry and Molecular Biology. While we will measure and plot the indicator values for non-fossil fuels, we will not include it in data fits and statistical analysis.   
 
 \begin{table}
 \centering
  \footnotesize
  \begin{adjustbox}{angle=90}
\begin{tabularx}{1.33\textwidth}{ c | c | c | c | c | }
\textbf{RET} & \textbf{science dependence} & \textbf{internal dependence} & \textbf{inter-patent distance in km} & \textbf{reference distance in km} \\
 & EP \quad \quad US & EP \quad \quad US & EP \quad \quad US & EP \quad \quad  US \\
\hline
\includegraphics[width=20pt]{geo.png} & 0.11(0.54) \quad 0.63(3.57) & 3.34(3.04) \quad 5.63(8.61) & 3284(3546) \quad 4844(3785) &  3803(2558) \quad 3117(1714) \\
  \hline
  \includegraphics[width=20pt]{hydro} & 0.10(0.60) \quad 0.11(2.66) & 3.19(2.73) \quad 3.14(6.94) & 3755(3730) \quad 5654(3732) &  4388(2858) \quad 3442(1786)\\
    \hline
  \includegraphics[width=20pt]{sea.png} & 0.19(1.93) \quad 0.52(3.21) & 3.96(3.54) \quad 6.86(10.51) & 3937(3689) \quad 5689(3677)  &  4375(2529)\quad 3502(1674)\\
    \hline
  \includegraphics[width=20pt]{sol.png} & 0.19(1.18) \quad 0.53(3.20) & 4.74(3.75) \quad 8.41(18.54) & 3953(4046) \quad 5500(3854) &  4085(2850)\quad 3427(1808)\\
    \hline
  \includegraphics[width=20pt]{photo.png} & 1.85(5.90) \quad 3.12(13.51) & 3.88(7.35) \quad 7.09(19.53) & 6023(3992) \quad 6285(3984) & 5332(2791)\quad 4185(2029)\\
    \hline
   \includegraphics[width=20pt]{wind.png} & 0.29(1.29) \quad 0.37(3.49) & 3.95(3.14) \quad 6.89(10.30) & 4308(3796) \quad 5439(3672) & 3451(2495)\quad 3923(1774)\\
    \hline
   \includegraphics[width=20pt]{cleancom.png} & 0.24(1.25) \quad 0.96(6.37) & 1.85(1.99) \quad 4.97(17.12) & 5392(3895) \quad 5957(3919) &  3867(2628)\quad 3683(1927)\\
    \hline
   \includegraphics[width=20pt]{nuc.png} & 0.26(1.18) \quad 0.66(2.18) & 3.06(2.40) \quad 4.18(7.70) & 5346(3677) \quad 5814(3774) &  4553(3201)\quad 3685(2471)\\
    \hline
   \includegraphics[width=20pt]{smart2.png} & 0.63(1.63) \quad 1.38(5.47) & 2.07(2.05) \quad 3.65(5.02) & 5089(3941) \quad 5958(3906) &  4526(2831)\quad 3678(2009)\\
    \hline
   \includegraphics[width=20pt]{bio.png} & 9.83(65.76) \quad 11.69(47.30) & 3.25(4.89) \quad 4.84(9.36) & 4527(3837) \quad 5529(3756) &   3671(2659)\quad 3475(1837) \\
    \hline
   \includegraphics[width=20pt]{stor.png} & 0.71(3.22) \quad 2.03(9.59) & 2.11(2.35) \quad 3.53(7.30) & 5501(4216) \quad 5579(4333) &  4846(2903)\quad 4257(2109)\\
    \hline
  \includegraphics[width=20pt]{hydrogen.png} & 1.22(4.52) \quad 2.38(11.80) & 2.08(2.41) \quad 3.59(6.08) & 5564(3920) \quad 6227(3887) &  4735(2684) \quad 3659(1909)\\
    \hline
 \includegraphics[width=20pt]{fuelcell.png} & 1.11(3.81) \quad 2.72(9.74) & 1.94(2.40) \quad 3.14(6.85) & 6176(3926) \quad 5945(4206) &  5587(2536)\quad 4130(1958)\\
    \hline
    \includegraphics[width=20pt]{ccs.png} & 1.01(4.75) \quad 3.44(13.05) & 2.52(2.71) \quad 5.75(10.97)  & 5578(3790) \quad  5732(3812) & 4265(2403)
 \quad 3610(1889) \\
    \hline
\end{tabularx}
\end{adjustbox}
\caption{\textbf{Descriptive statistics of main indicators} Note that all presented indicators are averages, the standard deviations are included in brackets. The units of the science and internal dependence are in reference/patent. All of the considered indicators are positive values only and highly skewed to zero. As explained earlier in Section \ref{indicators}, the reference distance is determined for a sub-selection of the patents.}
\label{RETS_des}
 \end{table}
 \begin{figure}
 \centering
  \includegraphics[width=\linewidth]{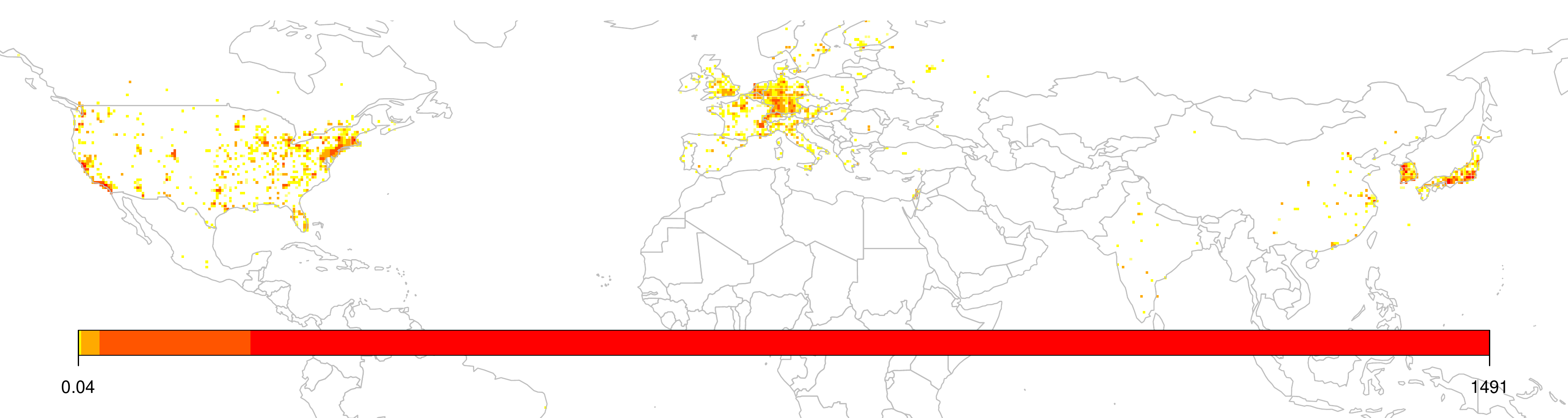}
  \caption{\textbf{Worldwide distribution of photovoltaics US patents based on inventor or applicant address} We plot the number of US patents  per grid-cell for a grid defined for each longitudinal and latitudinal half degree. The scale chosen for the color coding of the cells is logarithmic (see scale-legend).}
  \label{photo_world}
\end{figure}
\begin{figure}
\centering
  \includegraphics[width=\linewidth]{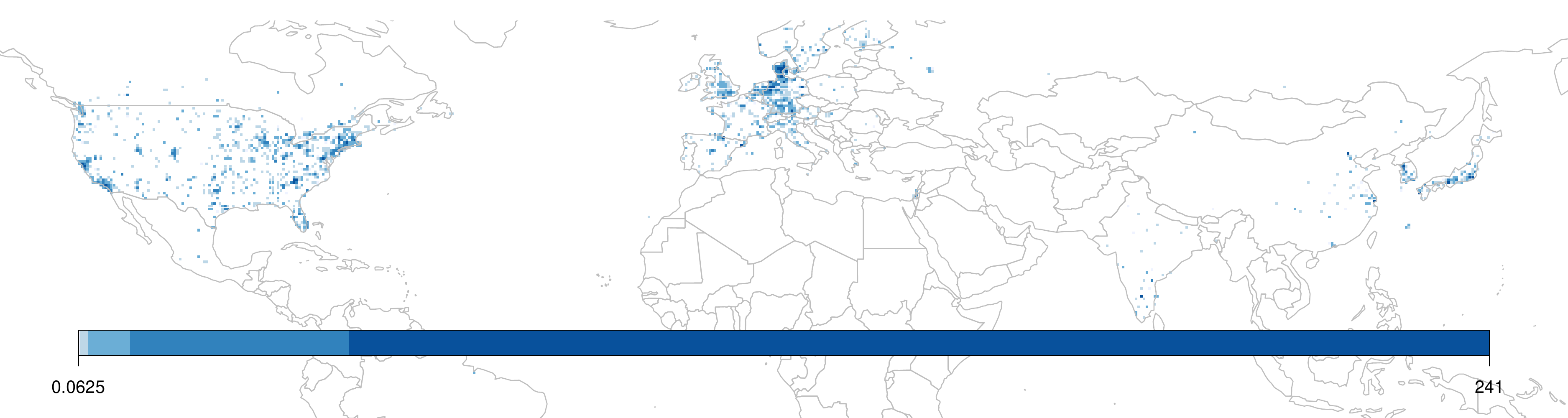}
  \caption{\textbf{Worldwide distribution of wind turbine US patents based on inventor or applicant address} Similar to Figure \ref{photo_world}, except for adjusting the color scale (the maximum is here 241 patents).}
  \label{wind_world}
\end{figure}
To explore the geographical distributions of inventive activity in RETs worldwide, we use the geographical coordinates to plot the number of US patents (using the inventor or applicant address in the patents) in a grid defined for each longitudinal and latitudinal half degree. We do this for photovoltaics and  wind-turbines respectively in Figures \ref{photo_world} and \ref{wind_world}. Because the patenting activity is distributed highly unevenly (a small number of areas producing the majority of patents), we chose a coloring following a logarithmic scale. We observe some variation between the figures, Germany and France innovating strongly in photovoltaics, while Denmark focusing more on wind-turbines. The main observation however, at least on a global scale, is that the geographical distributions of innovative activities are fairly similar, even for rather different technologies such as photovoltaics and wind turbines. 
In fact in a ranking of countries by total number of US patents, see appendix \ref{app_1}, the US, Japan and Germany are consistently in the top 5 for each RET considered in this research (and France for all but 3 RETs). For EP patents, these countries likewise dominate each top 5. Together these four countries account for 76 and 58  respective percentages of the US and EP patents (for RETs). This is line with the findings of earlier literature considering energy technology in general \cite{bointner_innovation_2014}. The uneven distribution is not due to our choice for counting at the country level. When instead consider spatial the level below countries (corresponding to the 'name\_1' level in the Geocoding database), we again see the same regions or locations recurring: California, New York, Tokyo, Bayern and Baden-W{\"u}rttemberg rather consistently dominate in the top 10 locations with most patents for each considered RET. An important part of the knowledge base development of RETs therefore appears to take place in a small number of dominant areas. Together with the similarity of the worldwide geographical distributions, these are relevant descriptive statistics: it indicates that despite the obvious location-boundness of \textit{the application} of specific RETs (hydro energy near rivers, photovoltaics in sunny locations, wind turbines near windy locations, etc.), the development of the knowledge base of these RETs still largely occurs in dominant areas which work on the development of all RETs at the same time.    

\section{Results}\label{results}

We will start this section by exploring the general relations between the analyticity, cumulativeness and the knowledge mobility, where we consider a general data set of patents. We then focus the analysis on the considered RETs, thereby discussing the various relations between indicators both qualitatively and quantitatively.  Following the first expectation in Section \ref{theory}, we expect to observe a positive relation between analyticity and knowledge mobility. Following the second expectation in Section \ref{theory}, we expect to see a negative relation between the cumulativeness and the knowledge mobility. 
\begin{figure}
\centering
\begin{minipage}{.48\linewidth}
  \includegraphics[width=\linewidth]{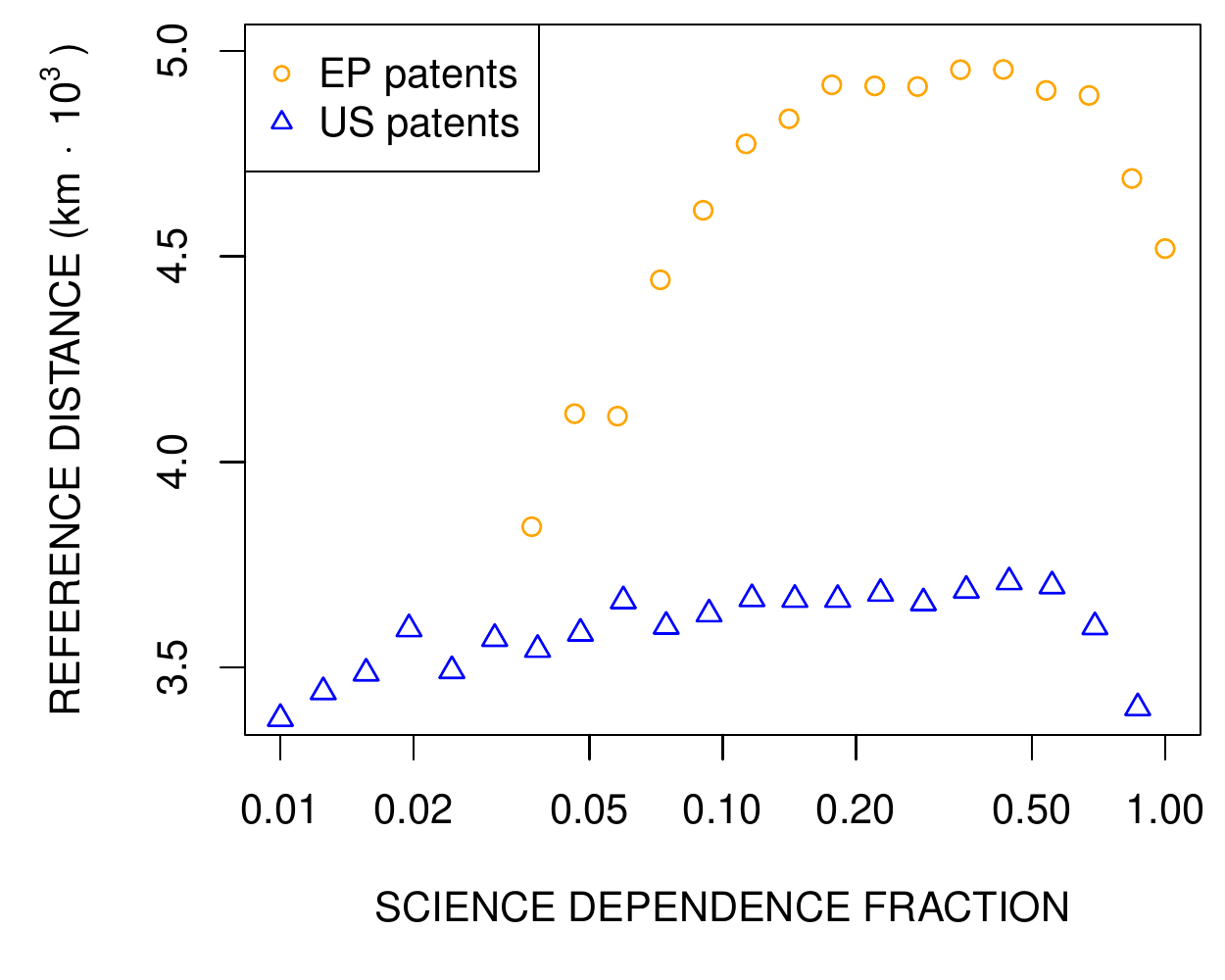}
\end{minipage}
\hspace{.02\linewidth}
\begin{minipage}{.48\linewidth}
  \includegraphics[width=\linewidth]{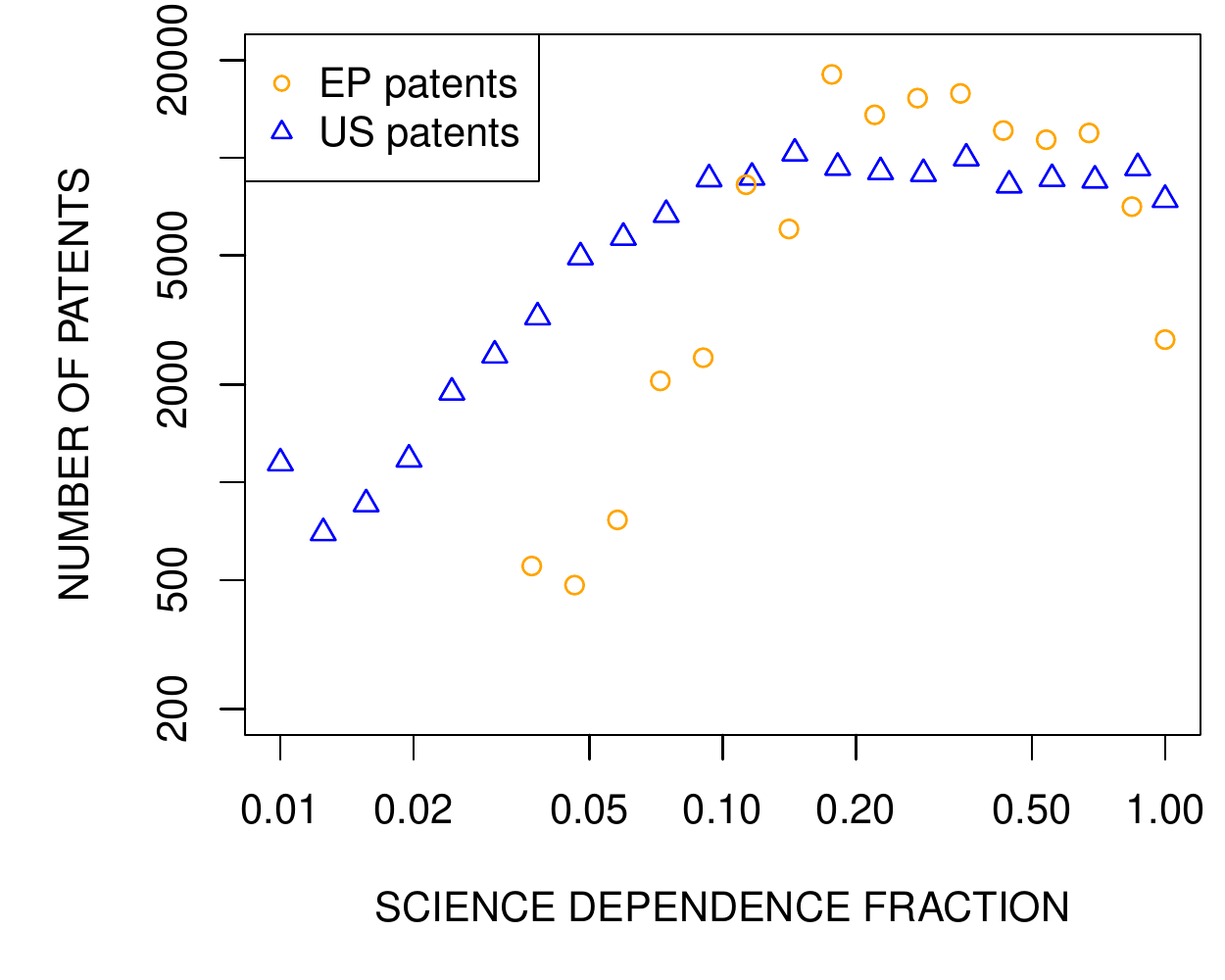}
\end{minipage}
  \caption{\textbf{Science dependence fraction, reference distance and number of patents} We divide the sdf into exponential bins (base 1.25) and determine for each bin the average rd (left panel) and the normalized cumulative number of patents (right panel). Note the sdf axes are logarithmic, hence the exponential bin sizes are constant in this plot. For sdf bins lower than 0.5  we observe a positive relation between the sdf and rd.}
  \label{sdf_general}
\end{figure}

\subsection{General relations between knowledge dimensions}

We will first explore some general relations between on the one hand the knowledge mobility and on the other hand analyticity or cumulativeness of knowledge. We explore these relations using the indicators that are defined on the level of individual patents: the science dependence fraction (sdf), the internal dependence fraction (idf) and the reference distance (rd). In the following analysis we include all EP and US patents for which a reference distance could be determined. One exception is the analysis of the US sdf: there we include, due to calculation challenges, a random selection of 20 percent of all such patents. As explained in Section \ref{indicators}, we sub-select those EP patents for which the inventors are from Europe and those US patents for which the inventor is from the US. To calculate the idf, the internal references are determined using within CPC-group references. 

In Figure \ref{sdf_general} we divide the sdf in exponential bins and plot the average rd (left panel) and number patents (right panel) for each bin. We observe in the left panel that the rd increases with increasing sdf for both EP and US patents (though more strongly for EP patents). Not included in this Figures are the many patents for which the sdf is zero ($7.0\cdot 10^5$ EP patents and $2.8\cdot 10^5$ US patents, respectively 6.4 and 2.0 times the total number of EP and US patents in Figure \ref{sdf_general}). The average rd of these zero sdf patents are 4114 km for EP patents and 3380 km for US patents, which are similar values to those in the lowest sdf bins in Figure \ref{sdf_general} and therefore in accordance with the observed relation. Even though the reference distance appears to go down for large sdf for both the EP and US patents, we note from the right panel that there relatively few patents with an sdf$>0.5$ (to be precise respectively 4 and 8 percent of the total EP and US patents). For the majority of the patents it therefore counts, in line with expectation, that the greater the sdf, the greater the rd. In other words, greater analyticity can be associated with greater knowledge mobility.
\begin{figure}
\centering
\begin{minipage}{.48\linewidth}
  \includegraphics[width=\linewidth]{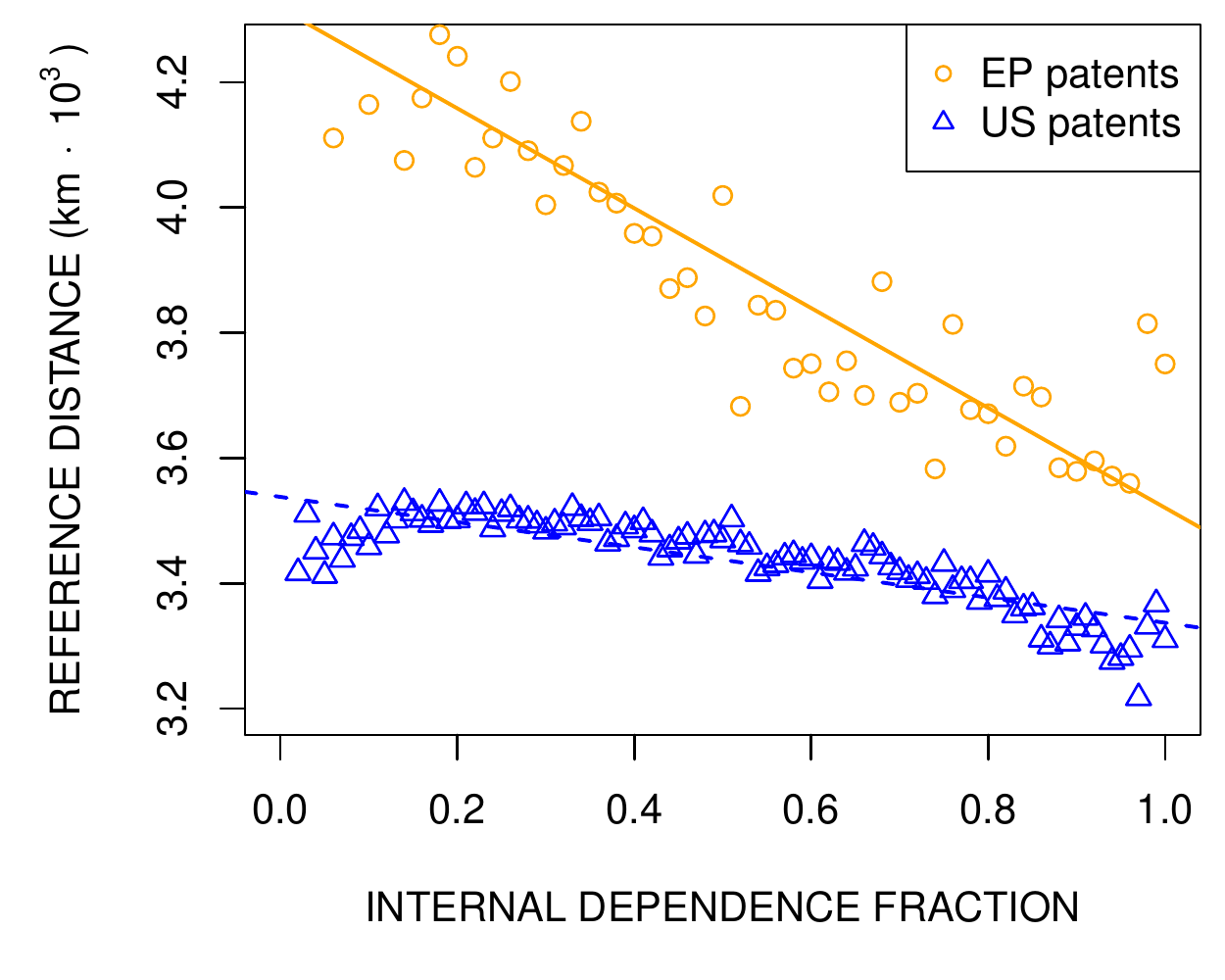}
\end{minipage}
\hspace{.02\linewidth}
\begin{minipage}{.48\linewidth}
  \includegraphics[width=\linewidth]{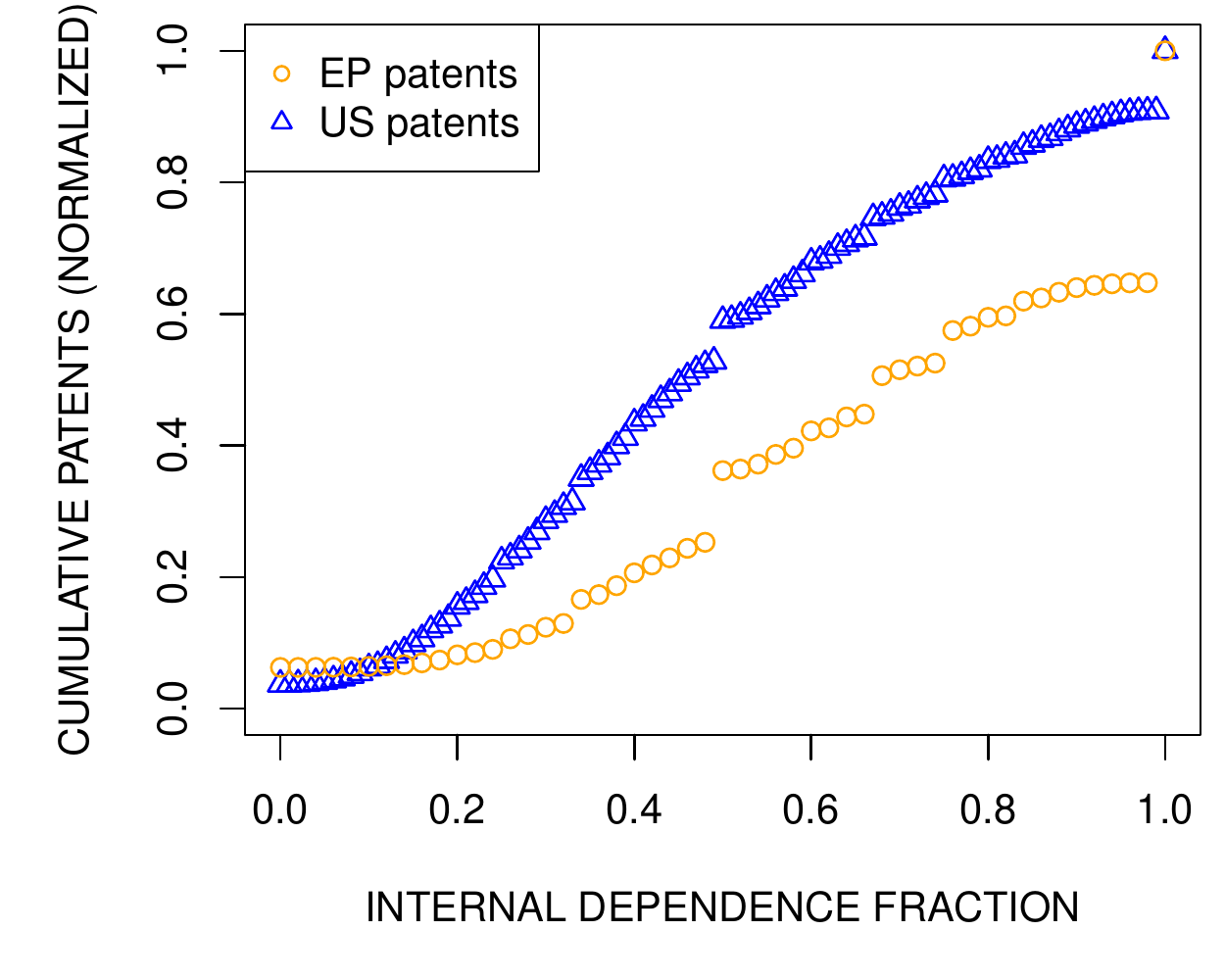}
\end{minipage}
  \caption{\textbf{Internal dependence fraction, reference distance and number of patents} We divide the idf into bins of constant size (0.01 for US and 0.02 for EP patents) and determine for each bin the average rd (left panel) and the normalized cumulative number of patents (right panel). For the rd linear fits are included, which indicate a negative relation between the idf and rd. Because the idf is a fraction, we observe small breaks in the right panel for highly frequent values such as $1/2$ and $2/3$.}
  \label{idf_general}
\end{figure}

In Figure \ref{idf_general} we divide the idf in bins of constant size  and plot the average rd (left panel) and cumulative number of patents (right panel) for each bin. We clearly observe that the rd decreases when the idf decreases (illustrated also by the linear fits). This is the case almost over the entire range of the idf. A minor exception are the US idf values $<0.15$. As is clear from the right panel however, there are relatively little patents in this range. As the right panel in Figure \ref{idf_general} illustrates, most of the patents have mid-range idf values, although there are relatively many patents with idf equal to one (counted in the bin with the largest idf value). As the left panel illustrates, the average rd of the patents in this bin are however in line with the observed pattern. We therefore conclude, in line with expectation, that the greater the idf, the smaller the rd. In other words, greater technological cumulativeness can be associated with lesser knowledge mobility

\subsection{Knowledge relations of RETs} 
Where in the previous section we discussed the general relations between knowledge dimensions based on a general data set of patents, we will in the following analyze these relations specifically considering the RETs. We will first qualitatively discuss the relation between on the one hand a knowledge mobility indicator and on the other hand either an analyticity or cumulativeness indicator. Subsequently, we will consider these relations more quantitatively, where we determine the correlations and estimate some models. 

\begin{figure}
\centering
\begin{minipage}{.48\linewidth}
  \includegraphics[width=\linewidth]{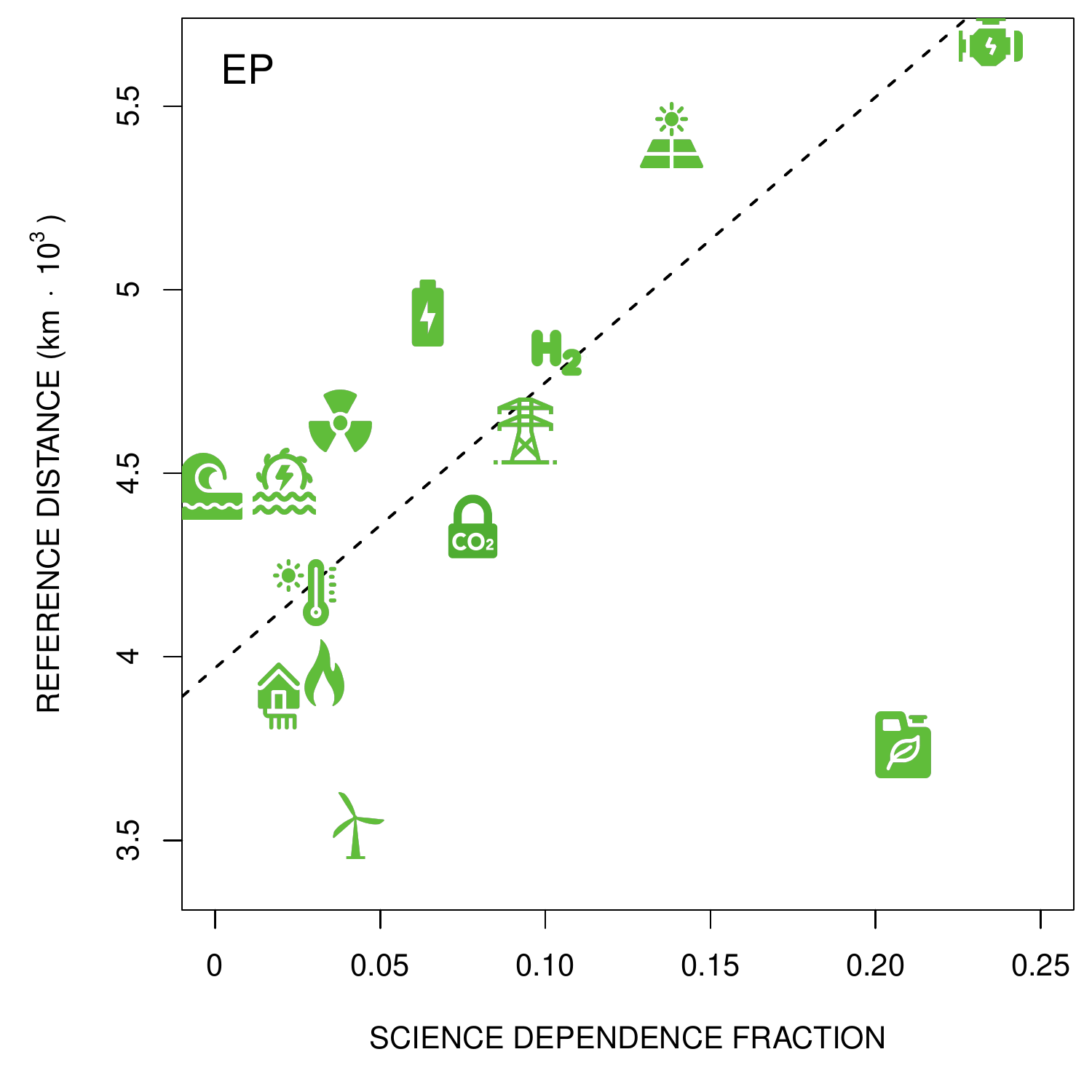}
\end{minipage}
\hspace{.02\linewidth}
\begin{minipage}{.48\linewidth}
  \includegraphics[width=\linewidth]{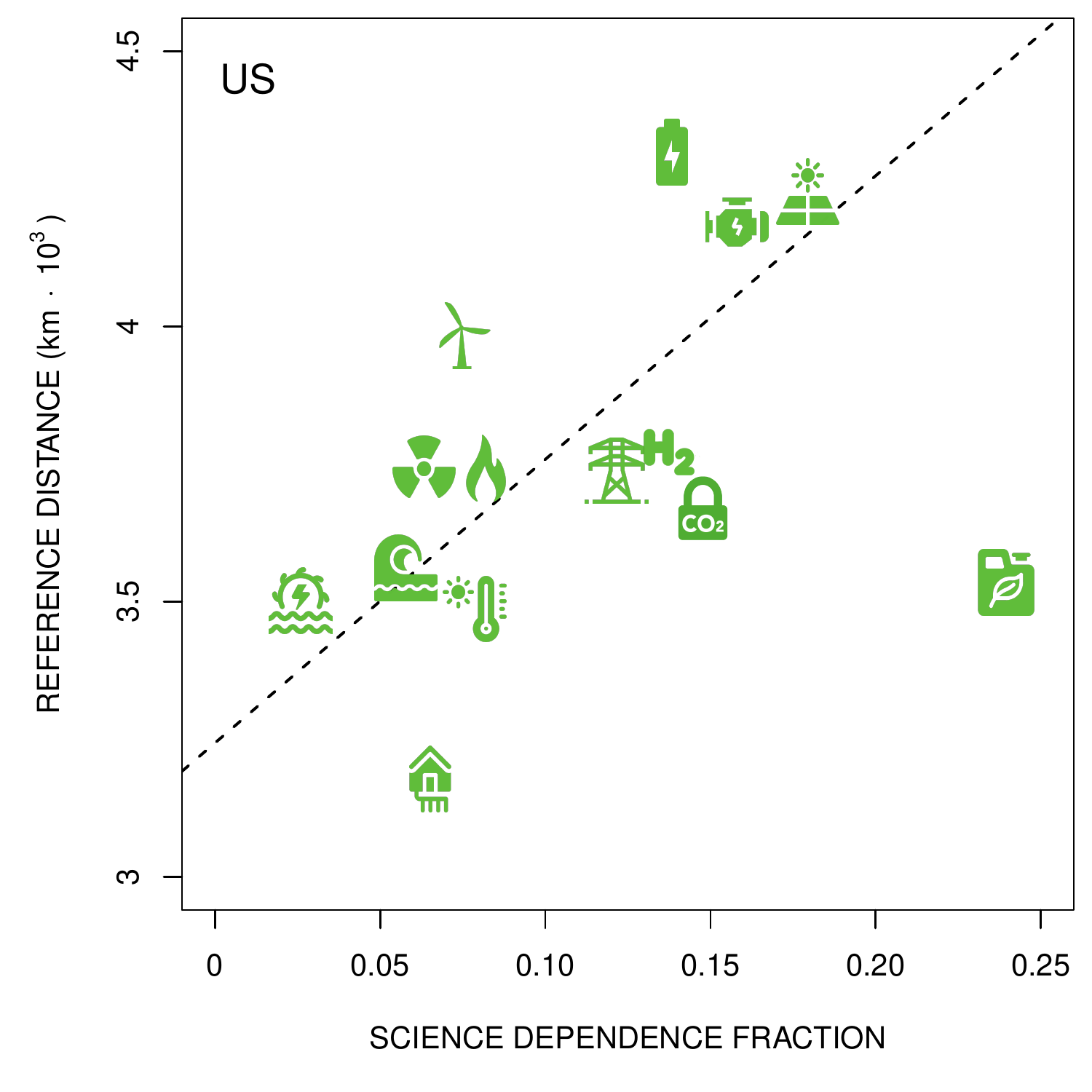}
\end{minipage}
\caption{\textbf{Reference distance for science dependence fraction} On the left we display EP patents and on the right US patents. See Table \ref{RETS} for a legend of the icons. Non-fossil fuels is excluded in the linear fit.}
\label{sci_dis}
\end{figure}

\begin{figure}
\centering
\begin{minipage}{.48\linewidth}
  \includegraphics[width=\linewidth]{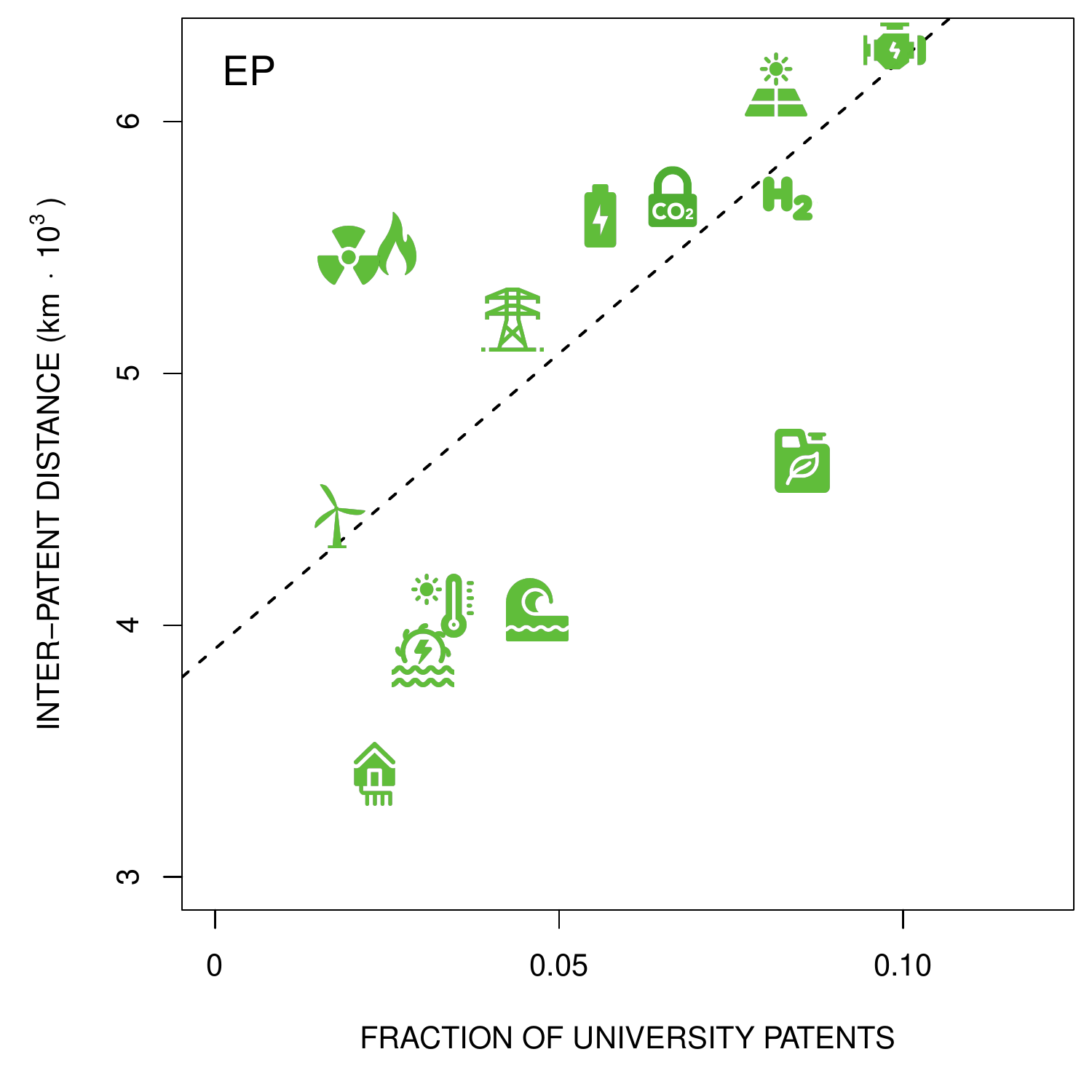}
\end{minipage}
\hspace{.02\linewidth}
\begin{minipage}{.48\linewidth}
  \includegraphics[width=\linewidth]{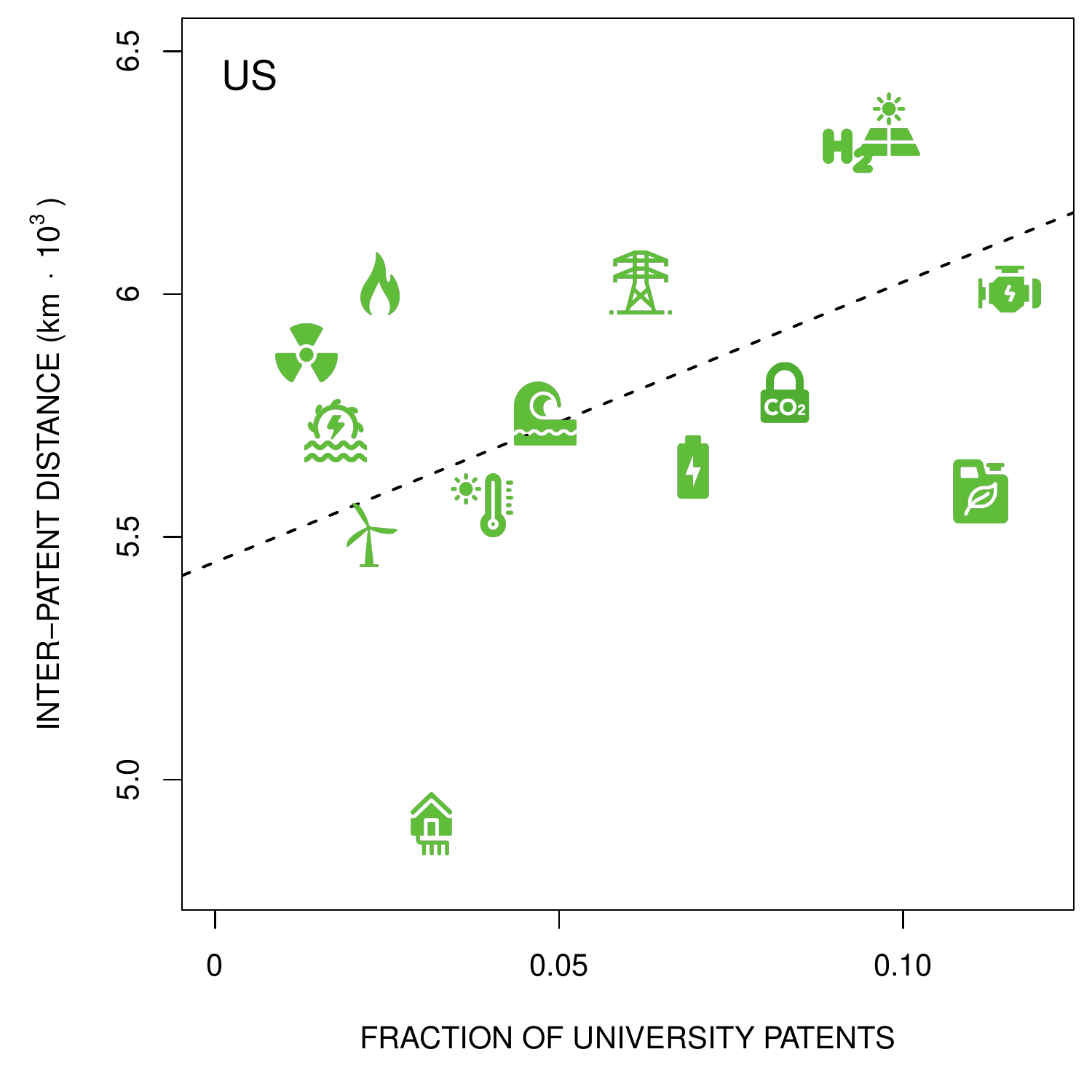}
\end{minipage}
  \caption{\textbf{Inter-patent distance for the fraction of university patents} On the left we display EP patents, on the right US patents.  See Table \ref{RETS} for a legend of the icons.}
  \label{uni_dis}
\end{figure}

In Figure \ref{sci_dis} we plot the rd for the sdf for both the EP patents (left) and the US patents (right). The main observation for both graphs is a positive relation between both quantities which is well fitted by a linear relation. We refer to Table \ref{RETS} for a legend of the icons and the short-names of the technologies. The sdf of non-fossil fuels can be observed to exceptionally large, which is, as discussed earlier, not included in these and later fits. It is therefore also challenging to compare the rd of this technology to the rest. It appears the values of the other technologies do allow for comparison however, and in line with expectation, technologies such as wind turbines, geothermal, hydro, solar thermal and energy from sea show relatively low rd, whereas photovoltaics, fuel-cells, energy storage and hydrogen technology show relatively large rd. Nuclear fission, ccs, clean combustion and electric grid technology are somewhat in between these two groups. Using Table \ref{RETS}, we note that technologies with a large number of patents (wind turbines, solar thermal, photovoltaics, fuel-cells) occur on both ends of the spectrum. It seems therefore that sheer numbers of patents, often a proxy for the size of the knowledge base, are not sufficient to explain the observed relation.    
In Figure \ref{uni_dis} we instead plot the inter-patent distance (ipd) for the university fraction (uf). The positive relation from Figure \ref{sci_dis} remains largely unchanged, the technologies we find upper right (down left) in Figure \ref{sci_dis} also tend to be in the upper right (down left) of Figure \ref{uni_dis}. This indicates that the variation across RETs in the considered knowledge dimensions is consistent for the different indicators for these dimensions. A closer look reveals some minor variations. The differences between the EP and US patents in Figure \ref{sci_dis} are relatively large in particular for wind turbines and fuel cells. In Figure \ref{uni_dis}, these differences are relatively less. This suggest that the ipd indicator may be more uniform between EP and US patents. We discuss the rd variations (and in particular those of wind turbines and fuel cells) in more detail in a part of Appendix \ref{app_2}. Especially for the US patents, the uf of nuclear fission and clean combustion is relatively low in Figure \ref{uni_dis} as compared to their sdf in from Figure \ref{sci_dis}. This suggests that the knowledge base of these technologies, while retaining a scientific component, is to a lesser extent developed in universities.   
While there are these minor differences, for most technologies the overall pattern is in agreement with Figure \ref{sci_dis}, again confirming that the technologies that build stronger on science also tend to show greater knowledge mobility.

\begin{figure}
\centering
\begin{minipage}{.48\linewidth}
  \includegraphics[width=\linewidth]{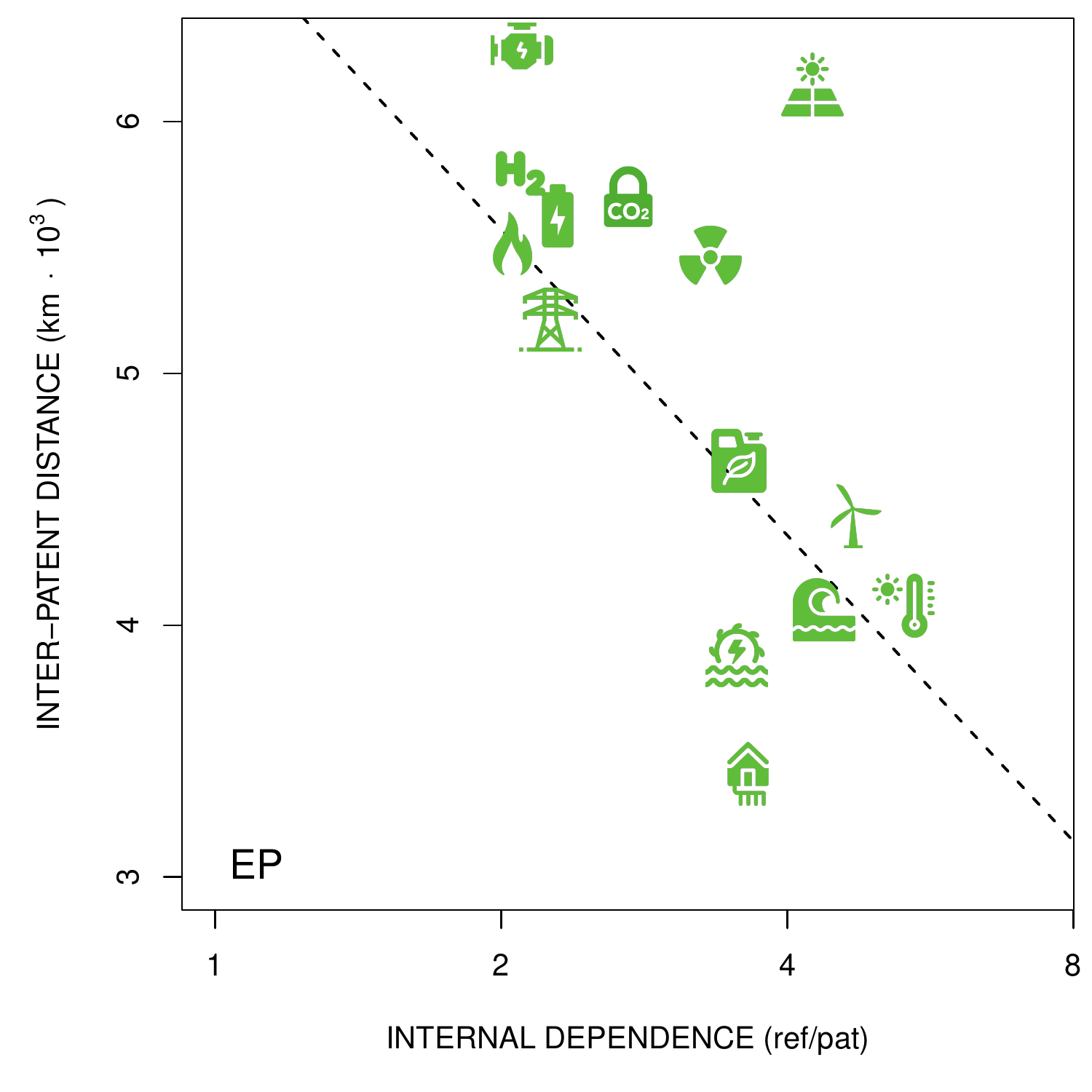}
\end{minipage}
\hspace{.02\linewidth}
\begin{minipage}{.48\linewidth}
  \includegraphics[width=\linewidth]{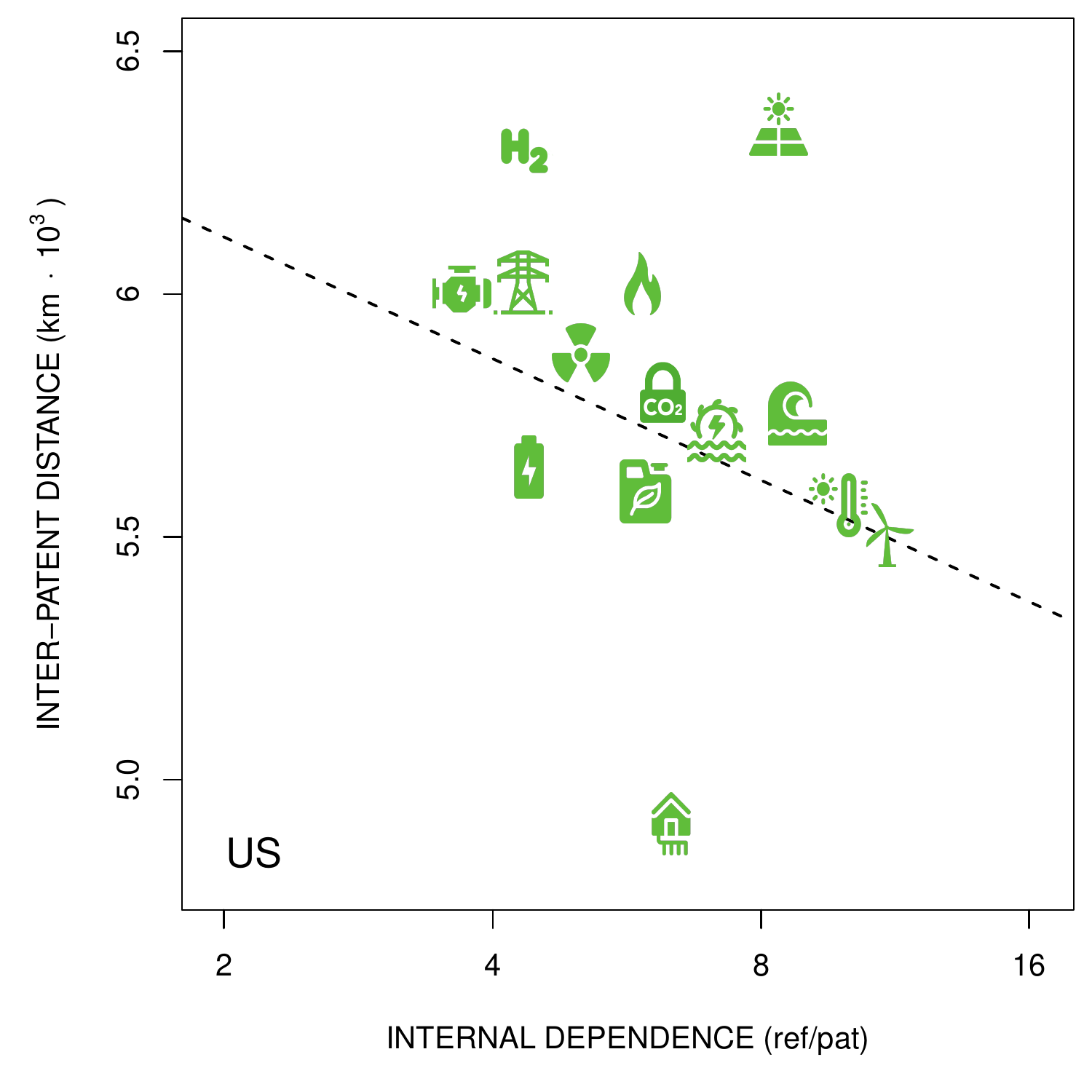}
\end{minipage}
  \caption{\textbf{Inter-patent distance for the internal dependence} On the left we display EP patents, on the right US patents. Note the horizontal axis is logarithmic. See Table \ref{RETS} for a legend of the icons.}
  \label{id_dis}
\end{figure}

\begin{figure}
\centering
\begin{minipage}{.48\linewidth}
  \includegraphics[width=\linewidth]{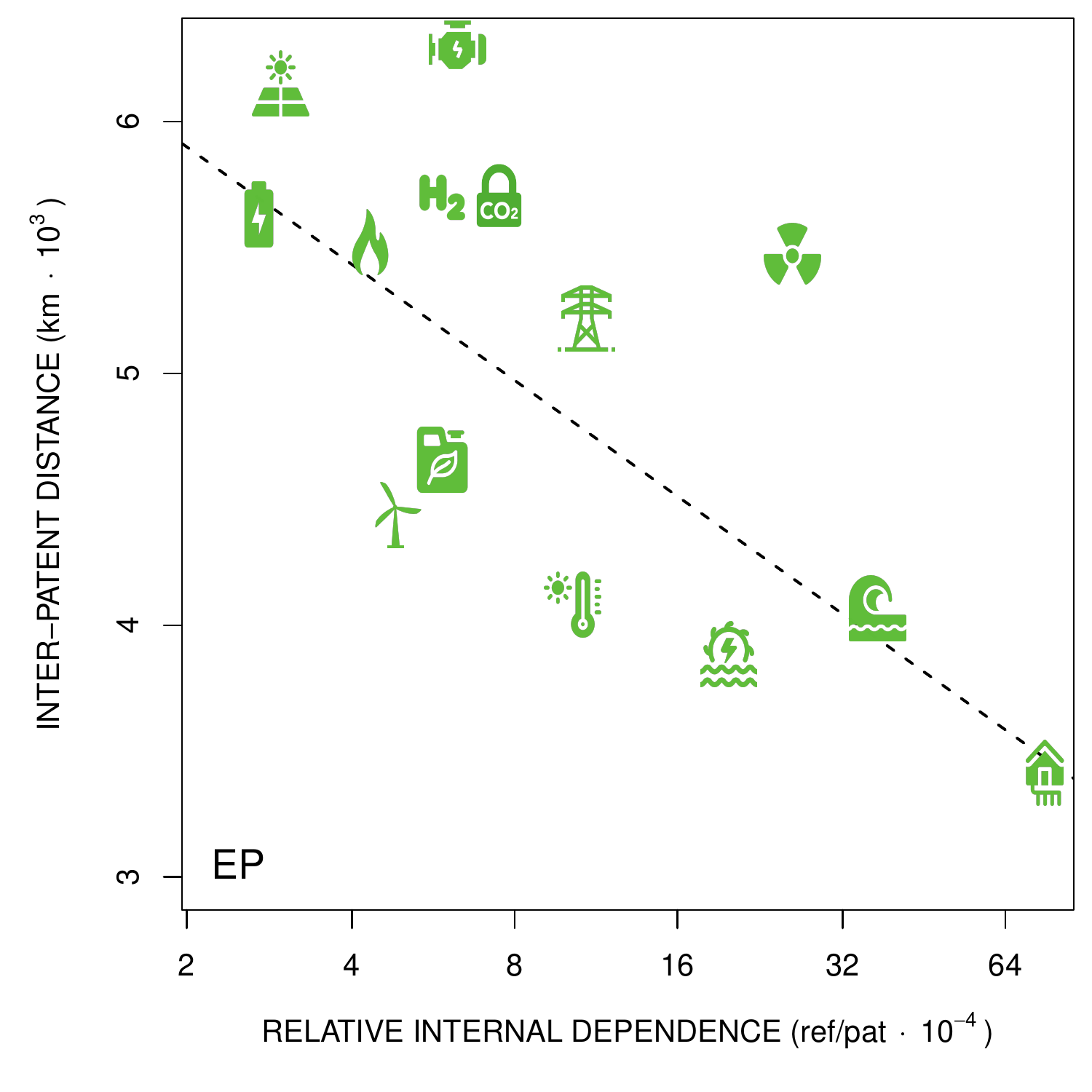}
\end{minipage}
\hspace{.02\linewidth}
\begin{minipage}{.48\linewidth}
  \includegraphics[width=\linewidth]{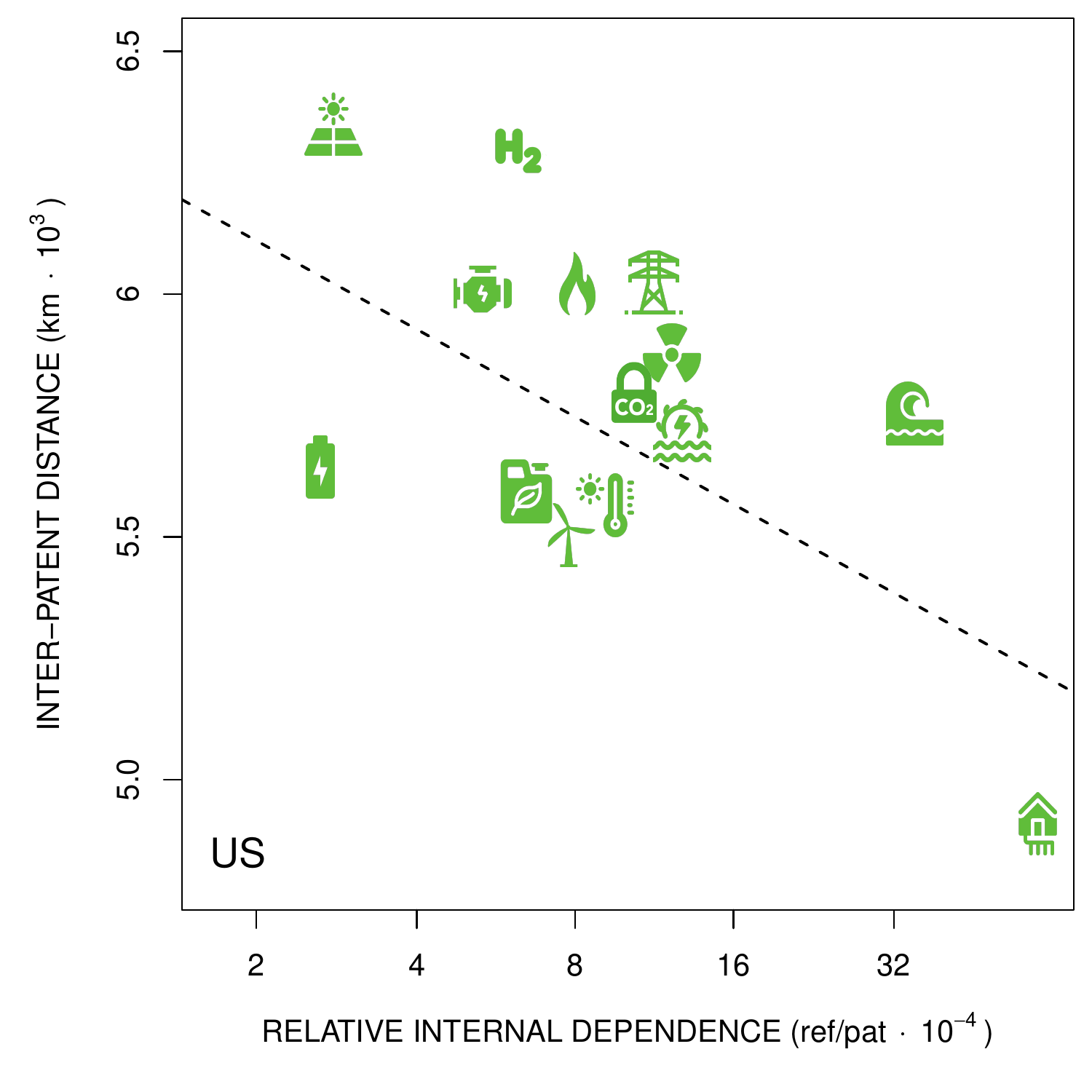}
\end{minipage}
  \caption{\textbf{Inter-patent distance for the relative internal dependence} On the left we display EP patents, on the right US patents. Note the horizontal axis is logarithmic. See Table \ref{RETS} for a legend of the icons.}
  \label{rid_dis}
\end{figure}

Next we plot the ipd for the internal dependence (id) in Figure \ref{id_dis}. The main observation is a negative relation between both quantities which is rather well fitted by a negative logarithmic relation (note the horizontal axis is logarithmic).\footnote{We may also take the log of the ipd and instead fit a power relation, the results will be largely comparable.} This is in line with the expected negative relation between knowledge mobility and technological cumulativeness. The only technology defying this pattern, both for EP and US patents but especially US patents, appears to be photovoltaics, which despite a relatively id, shows great ipd. In earlier contribution however we already demonstrated that the internal dependence tends to increase linearly with the number of patents \cite{persoon_how_2021}. Photovoltaics consists of far more patents than the other RETs (especially for US patents), which possibly explains the exceptional value for the internal dependence in this context.   
Alternatively, we may therefore consider the cumulativeness relative to the size of the knowledge base, which we measure by the relative internal dependence (rid) in Figure \ref{rid_dis}. In that figure the value of photovoltaics indeed shifts both for the EP and US patents to the left, in better agreement with its large value for the ipd. We observe a similar shift for wind turbines, though to a lesser extent, which is in line with expectation given its shorter ipd. Other than these changes the pattern is largely similar to the one in Figure \ref{id_dis}.  

\begin{figure}
\centering
\begin{minipage}{.48\linewidth}
  \includegraphics[width=\linewidth]{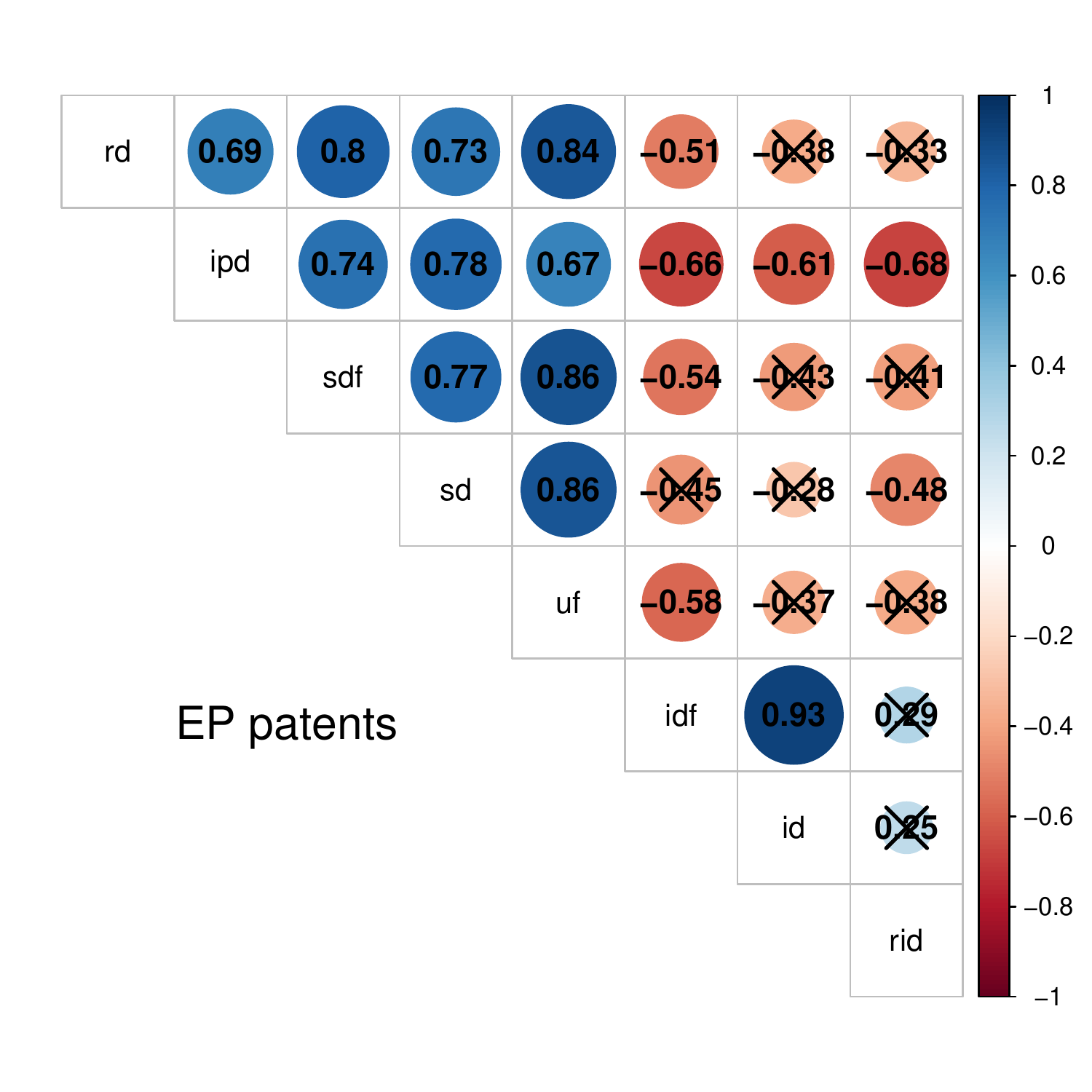}
\end{minipage}
\hspace{.02\linewidth}
\begin{minipage}{.48\linewidth}
  \includegraphics[width=\linewidth]{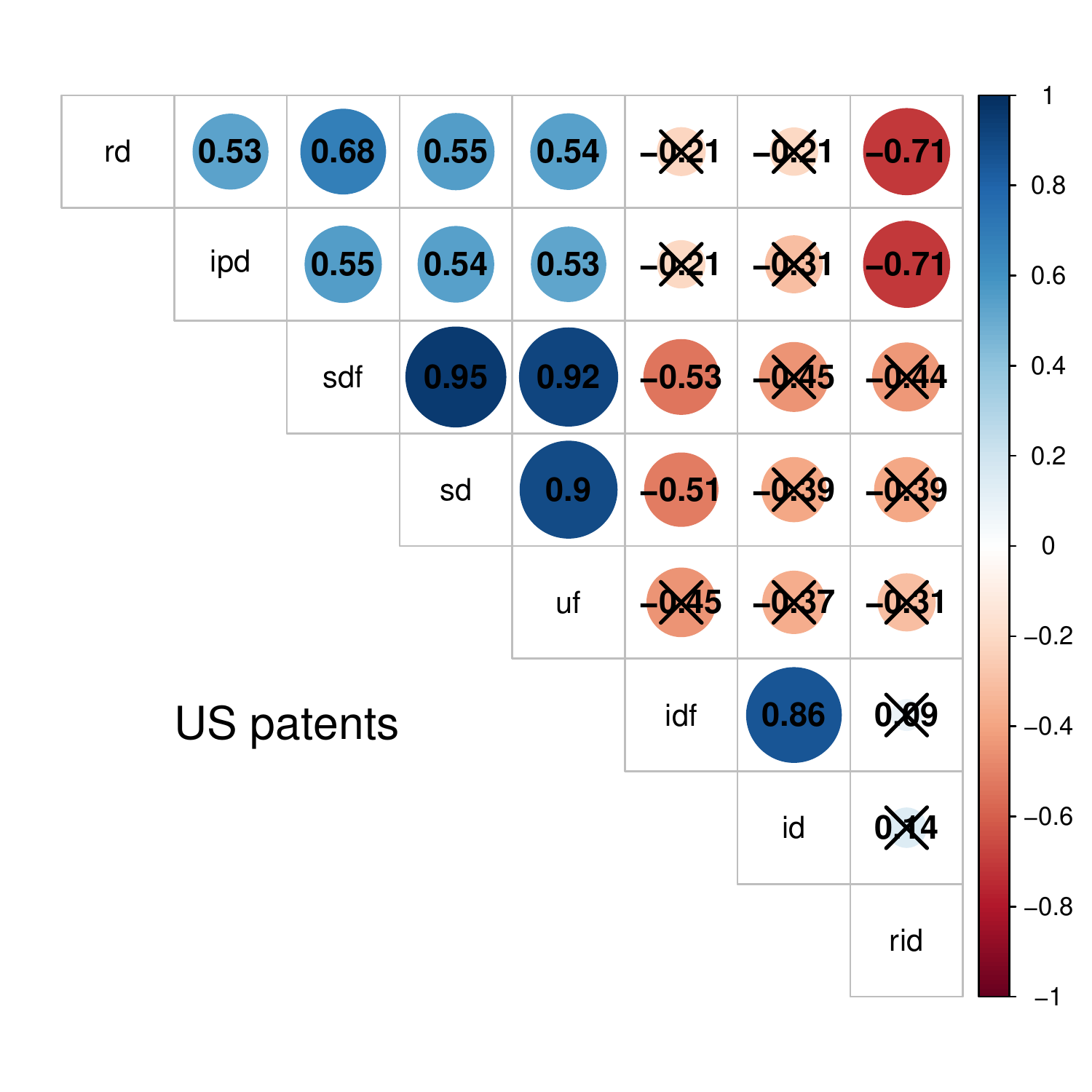}
\end{minipage}
  \caption{\textbf{Mutual correlations between indicators} On the left we display the correlations for EP patents, on the right for US patents. rd=reference distance, ipd=inter-patent distance, sdf=science dependence fraction, sd=science dependence,  uf=university fraction,idf= internal dependence,  id=internal dependence, rid=relative internal dependence. Each circle represents a mutual relation, the size and color of which represent the Pearson correlation coefficient. When a cross is included the relation is not significant on a 0.1 level. Non-fossil fuels are excluded while determining these correlations.}
  \label{mutual_correlations}
\end{figure}

Finally, note in both Figure \ref{id_dis} and \ref{rid_dis} that the values for non-fossil fuels are more or less in line with the rest of the technologies, where earlier for the science dependence its values were rather exceptional. This therefore presents an extra reason for considering both the science dependence and internal dependence: an exceptional value for the former need not automatically imply an exceptional value for the latter.  
We will not plot all possible combinations between the indicators we consider, yet for completeness we include in Figure \ref{mutual_correlations} the Pearson correlation coefficients of each combinations and whether or not this combination is statistically significant. We conclude from Figure \ref{mutual_correlations} that all correlation coefficients, most of which are substantial, have the expected sign: all analyticity indicators have positive signs with knowledge mobility indicators and all cumulativeness indicators have a negative sign with all knowledge mobility indicators. Especially the analyticity indicators show strong correlations with the knowledge mobility indicators. As expected the correlations between indicators of the same knowledge dimension are generally strong. One exception is the relative internal dependence (rid), which despite strong correlations with knowledge mobility indicators does not correlate strongly with other cumulativeness indicators. Interestingly, the rid does not (anti)correlate strongly with analyticity indicators either, which suggests its relation to the knowledge mobility is to some extent independent of the other indicators. We also observe this for internal dependence (id) of the European patents. To follow up on this suggestion, we will finally consider the possibility to model the knowledge mobility as a linear combination of an analyticity indicator and a cumulativeness indicator.
Again we will not present here all such possible linear combinations here in detail, there are simply too many, but instead share our general conclusions and include two examples (Table \ref{regression}). As we only consider 13 technologies, i.e. 13 data points, it does not make much sense go much further than combinations of 2 variables. We will first discuss this for the EP patents and than for the US patents.

When we model for the EP patents the internal patent dependence (ipd) as a linear combination of any given analyticity and any given cumulativeness indicator, the performance of the model in terms of minimizing the residual standard error and maximizing the Pearson correlation squared ($R^2$) is much better than for the case where all this indicators are individually considered as predictors. One specific example is given in Table \ref{regression}(left panel), where the EP ipd is modeled as a linear combination of the sdf and rid. The residual standard error (548.6) is much lower than that in a model with only the sdf (662) or the rid (724). The found $R^{2}=0.72$, corresponding to a Pearson coefficient of $R=0.85$, is also greater than the $R$ values in Figure \ref{mutual_correlations} for ipd-sdf and ipd-rid. When we take linear combinations of only analyticity or only cumulativeness indicators to model the ipd, this only results in a better model in half of the cases. This therefore indicates it makes sense to consider the analyticity and cumulativeness as independent factors relating to the knowledge mobility. As Figure \ref{mutual_correlations} already indicates, the EP patent reference distance very strongly correlates with most of the analyticity indicators, which is difficult to improve considering extra indicators. For the EP rd, we therefore only find very few combinations which present better models than the indicators considered individually. 

When we model the knowledge mobility indicators for the US patents as a linear combination of indicators we reach similar conclusions. We find for both knowledge mobility indicators that any combination between the rid and any analyticity indicator result in a better model than when the indicators are considered individually (again judged on the basis of the residual standard error and $R^{2}$). We present one example in Table \ref{regression} (right panel), where we model the US rd as a linear combination of the US sdf and US rid. The found residual standard error (207.4) is much smaller than that in a model with only the sdf (251) or only the rid (244). Also the found $R^{2}=0.67$, corresponding to $R=0.82$, is greater than the $R$ values in Figure \ref{mutual_correlations} for rd-sdf and rd-rid. We note that this is largely due to the success of the rid. It is not directly clear why this indicator, as compared to the other cumulativeness indicators, performs much better for the US patents. At least it underlines the need to consider multiple indicators to describe these knowledge dimensions. 
Only one combination (sd \& sdf) of either considering only analyticity indicators or only cumulativeness indicators can be evaluated as a better model than considering the indicators individually. This again indicates that especially the combination of a cumulativeness indicator and an analyticity indicator results in a better model, thus confirming the earlier assertion that the science and internal dependence are complementary indicators, both relating to the knowledge mobility.
\begin{table}
\centering
\footnotesize
\begin{minipage}{.48\linewidth}
\resizebox{\textwidth}{!}{%
\begin{tabular}{@{\extracolsep{5pt}}lc} 
\\[-1.8ex]\hline 
\hline \\[-1.8ex] 
 & \multicolumn{1}{c}{\textit{Dependent variable:}} \\ 
\cline{2-2} 
\\[-1.8ex] & ipd EP patents \\ 
\hline \\[-1.8ex] 
 sdf EP patents & 8,548$^{**}$ \\ 
  & (2,826) \\ 
  & \\ 
 rid EP patents & $-$225,936$^{**}$ \\ 
  & (91,980) \\ 
  & \\ 
 Constant & 4,726$^{***}$ \\ 
  & (302) \\ 
  & \\ 
\hline \\[-1.8ex] 
Observations & 13 \\ 
R$^{2}$ & 0.72 \\ 
Adjusted R$^{2}$ & 0.66 \\ 
Residual Std. Error & 548.6 (df = 10) \\ 
F Statistic & 12.8$^{***}$ (df = 2; 10) \\ 
\hline 
\hline \\[-1.8ex] 
\textit{Note:}  & \multicolumn{1}{r}{$^{*}$p$<$0.1; $^{**}$p$<$0.05; $^{***}$p$<$0.01} \\ 
\end{tabular} 
 
 }%
\end{minipage}
\hspace{.02\linewidth}
\begin{minipage}{.48\linewidth}
\resizebox{\textwidth}{!}{%
  \begin{tabular}{@{\extracolsep{5pt}}lc} 
\\[-1.8ex]\hline 
\hline \\[-1.8ex] 
 & \multicolumn{1}{c}{\textit{Dependent variable:}} \\ 
\cline{2-2} 
\\[-1.8ex] & rd US patents \\ 
\hline \\[-1.8ex] 
 sdf US patents & 3,490$^{**}$ \\ 
  & (1,521) \\ 
  & \\ 
 rid US patents & $-$112,047$^{**}$ \\ 
  & (45,023) \\ 
  & \\ 
 Constant & 3,535$^{***}$ \\ 
  & (180) \\ 
  & \\ 
\hline \\[-1.8ex] 
Observations & 13 \\ 
R$^{2}$ & 0.67 \\ 
Adjusted R$^{2}$ & 0.61 \\ 
Residual Std. Error & 207.4 (df = 10) \\ 
F Statistic & 10.2$^{***}$ (df = 2; 10) \\ 
\hline 
\hline \\[-1.8ex] 
\textit{Note:}  & \multicolumn{1}{r}{$^{*}$p$<$0.1; $^{**}$p$<$0.05; $^{***}$p$<$0.01} \\ 
\end{tabular} 

 }%
\end{minipage}
  \caption{\textbf{Examples of regression outcomes for linear models} We present the results of two regressions, one for EP patents (left panel) and one for US patents (right panel). On the left, we choose the ipd as the dependent variable and the sdf and rid as independent variables, where we also allow a constant term. On the right, we choose the rd as the dependent variable instead. Note there are 13 data points as we exclude non-fossil fuels.}
  \label{regression}
\end{table}

In sum, while there is considerable variation across different RETs in terms of knowledge mobility, this variation is to some extent explained by their variation in analyticity and cumulativeness, thus in line with the expectations of Section \ref{theory}. We can distinguish rather consistently a collection of footloose RETs (photovoltaics, energy storage and fuel cells) characterized by relatively high analyticity and low cumulativeness, from a collection of sticky RETs (energy from sea, wind-turbines, geothermal, hydro, and solar thermal energy) characterized by relatively low analyticity and high cumulativeness. 

\section{Discussion}\label{discussion}

In this research we established a close relationship between the analyticity, cumulativeness and the knowledge mobility of technology in general and in particular for various RETs. In this section we discuss several deeper theoretical aspects and  limitations of our approach. 

First, our results suggest that analyticity and cumulativeness are two distinct characteristics of a knowledge base, in same way that technological cumulativeness and building on scientific knowledge are two distinct mechanisms for technological change. While both the science dependence and internal dependence strongly relate to the knowledge mobility, we find that they are largely independent indicators, i.e., that a high value for the one need not imply a low value for the other. A comprehensive approach to the mechanisms underlying knowledge mobility therefore should not be limited to analyticity or technological cumulativeness but should instead treat these as complementary. 

Second, we emphasize our focus is on \textit{technological} cumulativeness in this research, i.e. studying the relevance of technological knowledge to later technological knowledge. This is not to mean scientific knowledge is not cumulative: 'scientific cumulativeness' should however be studied in the context of science building on science. Neither does it imply technology does not influence science: where technology provides science with the necessary instruments, science provides technology with the necessary analytical knowledge. 

We can also identify a number of limitations to our research. Our focus in this contribution is on the knowledge aspects of technology, though we acknowledge that there may be many more factors determining the geographical development of a technology, perhaps most importantly the (prospective) market valuation of that technology. For a more inclusive perspective we refer to \cite{binz_global_2017}. Furthermore, earlier contributions have argued that, while geographical distance remains an important or possibly the most important metric to measure knowledge mobility \cite{caragliu_space_2016}, a more comprehensive approach additionally includes a number of other metrics, based on, for example, organisational, institutional or cognitive proximity \cite{boschma_proximity_2005}. Recognizing this criticism, we performed additional analyses with alternative distance measures, such as the fraction of references staying with a region or country and the Herfindahl index of the distribution of patents over regions and countries. In both cases however the results were challenging to interpret, especially since we only considered 13 technologies. Where the fraction of references within region or country suggested contrary results for regions and countries, the Herfindahl index showed contrary results for EP and US patents (and showed some scaling with the number of patents, which further complicated matters). To keep this contribution simple, we excluded a detailed discussion of these results.  

\section{Conclusions and policy implications}\label{conclusions}

This paper contributes to the literature on local and global innovation systems through a systematic empirical analysis of the extent to which Renewable Energy Technologies (RETs) can be characterized as sticky or footloose \cite{binz_global_2017}. It illustrates the relation between the spatial innovation dynamics of technologies and characteristics of the knowledge base of these technologies, such as the extent to which this knowledge base is analytic (the 'analyticity') and the extent to which it is cumulative (the 'cumulativeness'). The tendency of technology to be spatially sticky or footloose can be systematically approached using concept of knowledge mobility, that is, the extent to which knowledge travels geographically. After empirically confirming, for general technology, the positive relation between analyticity and knowledge mobility and the negative relation between cumulativeness and knowledge mobility, we investigate these relations in more detail for various RETs. We find, in line with theoretical expectations, that the RETs with high analyticity, low cumulativeness knowledge bases (photovoltaics, fuel-cells, energy storage and hydrogen technology) show a greater knowledge mobility than those with low analyticity, high cumulativeness knowledge bases (wind turbines, geothermal, solar thermal, hydro energy and energy from sea). We will refer to the former group with 'analytic RETs' and the latter group with 'cumulative RETs'. Comparing non-fossil fuels to the other RETs is challenging, as its dependence on analytic knowledge appears to be exceptionally strong.  

Our findings lead to a number of recommendations for decarbonizing strategies and policies. For the transition from general R\&D stimulating and technology neutral subsidy schemes to more mission oriented science and technology policies, a deep understanding of the knowledge characteristics of the considered technology is key. As RET in general depends strongly on analytic knowledge, stimulating scientific research appears to be an effective and targeted measure to stimulate RET development. However, in this work we have demonstrated that there is also substantial variation across different RETs in various knowledge dimensions, and that this variation across RETs can be used to more effectively target the development of these RETs. More precisely, we have demonstrated that we can distinguish between analytic and cumulative RETs. Where the development of the former allows for easier entry and more flexibility in choosing locations, the development of the latter may be relatively harder to enter and is limited to locations providing the necessary synthetic knowledge. To encourage the development of analytic RETs in particular, policy makers may focus more on strengthening scientific activity. To encourage the development of cumulative RETs in particular, policy mixes focusing on system building are needed to stimulate the local presence of synthetic knowledge. In sum, our results call for policies which are more RET specific, taking into account the variation across RETs in various knowledge dimensions, which relate predictably to spatial dynamics of innovation. 

\section{Acknowledgements}

This work was supported by NWO (Dutch Research Council) grant nr. 452-13-010. The icons in Tables \ref{RETS} and \ref{RETS_des} and Figures 2-7 are made by Freepik, Kiranshastry, Pixel perfect and Pixelmeetup from www.flaticon.com.

\bibliographystyle{acm}
  \bibliography{Stickyness.bib}
 
 \newpage
\appendix
\section{Country ranking by number of patents}\label{app_1}

 \begin{table}[htp]
  \footnotesize
\begin{tabularx}{\linewidth}{ c|X|X|X|X|X|X|X|X|X|X|X|X|X|X|}
& \includegraphics[width=20pt]{geo.png} & 
  \includegraphics[width=20pt]{hydro} & 
  \includegraphics[width=20pt]{sea.png} &  
  \includegraphics[width=20pt]{sol.png} & 
  \includegraphics[width=20pt]{photo.png} & 
   \includegraphics[width=20pt]{wind.png} & 
   \includegraphics[width=20pt]{cleancom.png} & 
   \includegraphics[width=20pt]{nuc.png} & 
   \includegraphics[width=20pt]{smart2.png} & 
   \includegraphics[width=20pt]{bio.png} &
   \includegraphics[width=20pt]{stor.png} & 
   \includegraphics[width=20pt]{hydrogen.png} & 
    \includegraphics[width=20pt]{fuelcell.png} \\ 
    \hline
  1 & DE 100 & DE 236 & US 131 & DE	1039 & US 2129 & DE 1965 & US 1051 & US 370 & DE	386 & US 998 & JP 1221 & US	729 & US	623 \\
   \hline
    2 & US 45 & FR 146 & GB	98 & US	525 & JP 1857 & US 1058 & DE 886 & FR	216 & US	352 & DE 794 & DE 1009 & DE	487 & JP	497 \\
    \hline
     3 & CH	26 & US	132 & DE 82 & FR 338 & DE 1833 & DK 1003 & JP 485 & DE	145 & JP	243 & FR 373 & US 953 & JP 383 & DE	312 \\
     \hline
      4 & SE 19 & GB 11 & FR 69 & IT 255 & KR 848 & JP	590 & FR 315 & JP 115 & FR	111 & JP 164 & KR 508 & FR	347 & FR	139 \\
      \hline
       5 & FR 19 & JP 72 & NO 55 & ES 198 & FR 692 & ES	451 & CH 182 & SE 62 & SE	98 & NL	163 & FR 481 & IT 101 & KR	133 \\
    \hline
\end{tabularx}
\caption{\textbf{Country rankings of EP RET patents} We denote the countries by their alpha-2 letter codes and include the number of EP patents.}
\label{rankings_EP}
 \end{table}

 \begin{table}[htp]
  \footnotesize
\begin{tabularx}{\linewidth}{ c|X|X|X|X|X|X|X|X|X|X|X|X|X|X|}
& \includegraphics[width=20pt]{geo.png} & 
  \includegraphics[width=20pt]{hydro} & 
  \includegraphics[width=20pt]{sea.png} &  
  \includegraphics[width=20pt]{sol.png} & 
  \includegraphics[width=20pt]{photo.png} & 
   \includegraphics[width=20pt]{wind.png} & 
   \includegraphics[width=20pt]{cleancom.png} & 
   \includegraphics[width=20pt]{nuc.png} & 
   \includegraphics[width=20pt]{smart2.png} & 
   \includegraphics[width=20pt]{bio.png} &
   \includegraphics[width=20pt]{stor.png} & 
   \includegraphics[width=20pt]{hydrogen.png} & 
    \includegraphics[width=20pt]{fuelcell.png} \\ 
    \hline
   1 & US 450 & US 925 & US 700 & US 3921 & US 8154 & US 3759 & US 2801 & US 984 & US 1064 & US 3037 & US 3450 & US 2061 & US 1890 \\
   \hline
    2 & JP 38 & JP 159 & GB 91 & DE 488 & JP 4155 & DE 1501 & JP 740 & JP 295 & JP 390 & DE 395 & JP 2894 & JP 892 & JP 1291 \\
    \hline
     3 & DE 34 & DE 135 & FR 63 & JP 367 & KR 2318 & DK 858 & DE 548 & FR 196 & DE 302 & JP 326 & KR 1233 & DE 492 & KR 591 \\
     \hline
      4 & CA 33 & CA 97 & JP 63 & ES 167 & DE 1679 & JP 637 & FR 243 & DE 155 & KR 111 & FR 282 & DE 788 & FR 299 & DE 312 \\
      \hline
       5 & IL 21 & FR 95 & DE 47 & FR 166 & FR 627 & ES 392 & CH 153 & SE 71 & SE 93 & CA 200 & FR 378 & KR 200 & FR 149 \\
    \hline
\end{tabularx}
\caption{\textbf{Country rankings of US RET patents} We denote the countries by their alpha-2 letter codes and include the number of US patents.} \label{rankings_US}
 \end{table}
 
 \section{Reference distances of inventors in and outside Europe and US}\label{app_2}
 
\begin{figure}
\centering
\begin{minipage}{.48\linewidth}
  \includegraphics[width=\linewidth]{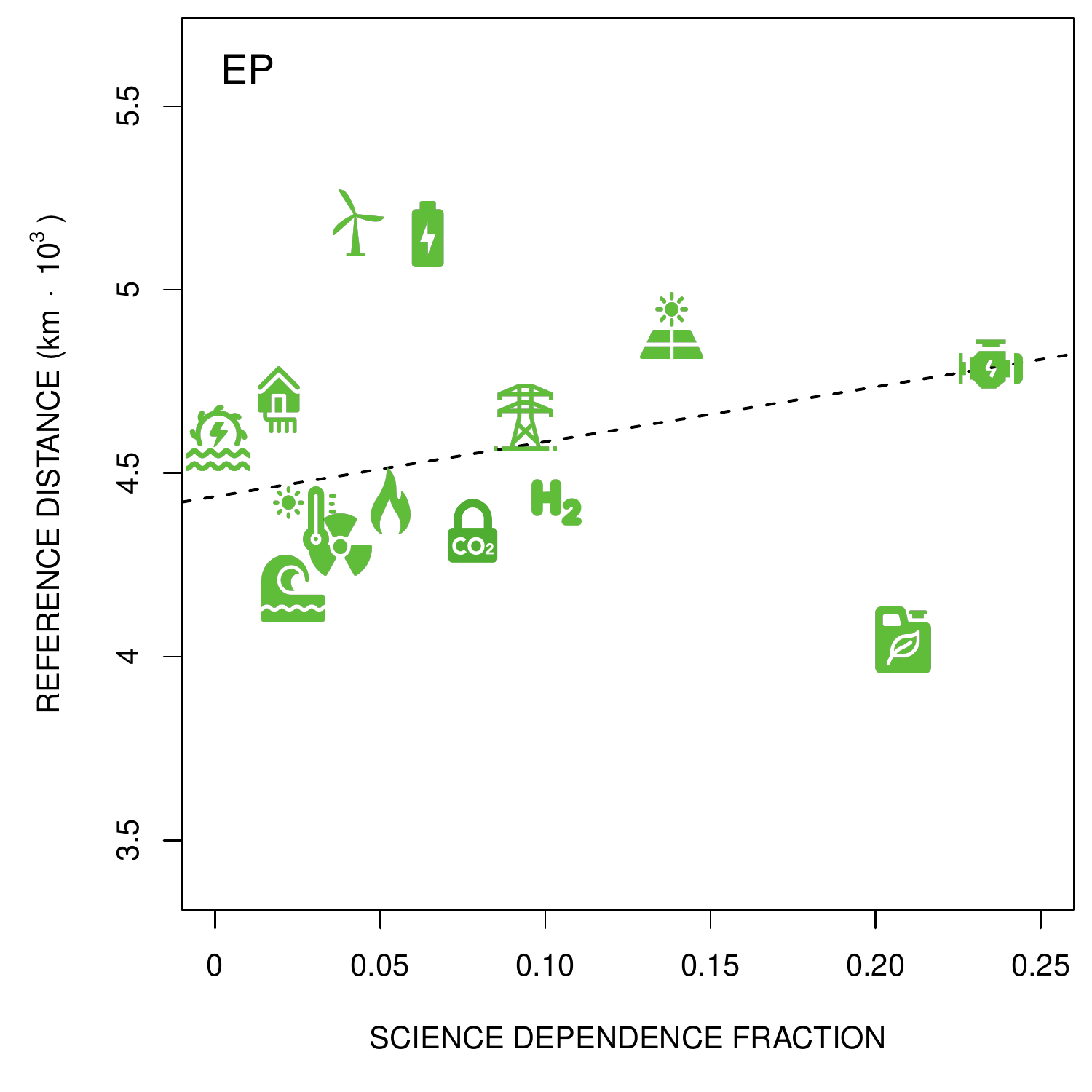}
\end{minipage}
\hspace{.02\linewidth}
\begin{minipage}{.48\linewidth}
  \includegraphics[width=\linewidth]{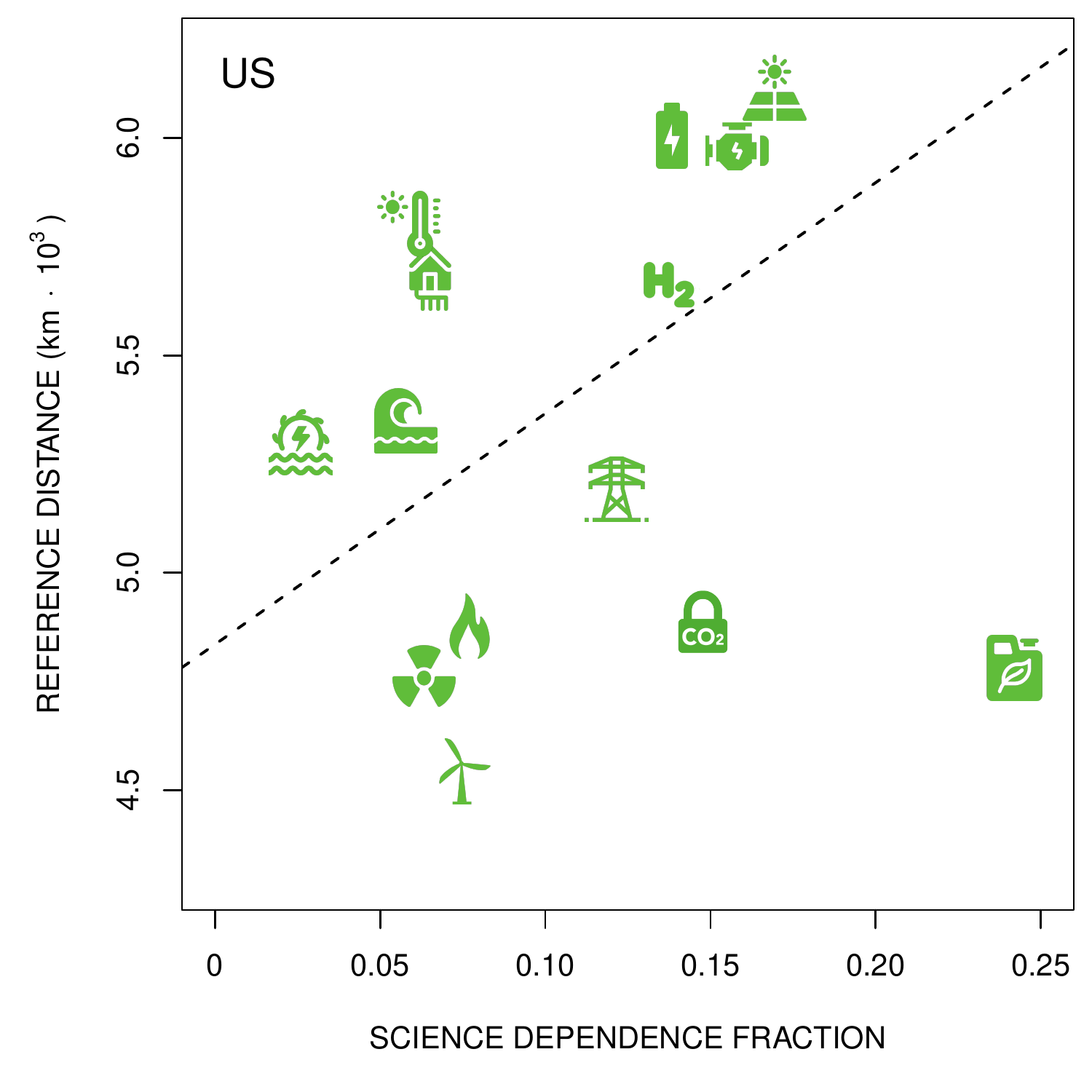}
\end{minipage}
\caption{\textbf{Reference distance for science dependence fraction} On the left we display EP patents of which the inventor is from the US and on the right we display US patents of which the inventor is from EP. See Table \ref{RETS} for a legend of the icons.}
\label{sci_dis_reversed}
\end{figure}

 In this appendix we discuss the effect of considering the reference distance of patents where the inventor is located (far) away from the jurisdiction of the patent office, for example EP patents with a US inventor (Figure \ref{sci_dis_reversed} left panel) or US patents with a EP inventor (Figure \ref{sci_dis_reversed} right panel).
 
In Figure \ref{sci_dis_reversed}, the relation between the science dependence fraction and the reference distance again appears to be positive, be it more irregular than the earlier observed relations. However, there are also a number of differences with Figure \ref{sci_dis}.
The most striking difference is that the reference distance of the US patents in \ref{sci_dis_reversed} are much greater. The reference distances can therefore be concluded to partly depend on the location of the inventor. If we would have included the US patents from Figure \ref{sci_dis_reversed} in Figure \ref{sci_dis}, this would especially affect the reference distance (which is an average over all patents) of technologies with lower numbers of patents. Another striking difference is that the positive relation between science dependence fraction and reference distance is for the EP patents a lot steeper in Figure \ref{sci_dis} than in Figure \ref{sci_dis_reversed}. This might be a result of the following. There are generally more US patents and inventors than EP patents and as a consequence the US patents are cited relatively often. The reference distances of an EP inventor referring to a US inventor is much larger than that of US inventor referring to a US inventor. There may therefore be less variation in the reference distance of US inventors, because even when they apply for a EP patent (i.e. Figure \ref{sci_dis_reversed} left panel), they may still be citing US inventors relatively often. Finally we note some typical differences on the level of individual technologies. Where wind turbines has a relatively high reference distance in the left panel of Figure \ref{sci_dis_reversed} (as compared to both the right panel and Figure \ref{sci_dis}). This indicates that the EP wind turbine inventors refer to patents from inventors close to home (most likely within Europe) whereas the US wind turbine inventors tend to refer to patents from inventors far from home (most likely Europe). This may illustrate a European lead in the wind turbine innovative activities. We see the reverse relation with fuel cells, of which the reference distance in the left panel of Figure \ref{sci_dis_reversed} is relatively low as compared to Figure \ref{sci_dis}. In the right panel its the reference distance is actually relatively large, indicating a US innovative lead for this technology.

\end{document}